\documentclass{jfm}
\pdfoutput=1
\usepackage{graphicx}
\usepackage{amsmath}
\usepackage{mathtools}
\usepackage{dcolumn}
\usepackage{bm}
\usepackage{multirow}
\usepackage{caption}
\usepackage{subcaption}
\usepackage{float}
\usepackage[dvipsnames]{xcolor}
\usepackage{tikz}
\usetikzlibrary{shapes}
\usetikzlibrary{plotmarks}
\usepackage{pgfplots}
\usepackage{comment}
\usepackage{hyperref}
\usepackage[symbol]{footmisc}

\usepackage{threeparttable}
\usepackage{lscape}

\newcommand{\blackline}{\raisebox{2pt}{\tikz{\draw[-,black,solid,line width = 0.75pt](0.,0mm) -- (5mm,0mm)}}}
\newcommand{\blackdashedline}{\raisebox{2pt}{\tikz{\draw[-,black,dashed,line width = 0.75pt](0.,0mm) -- (5mm,0mm)}}}
\newcommand{\blackdottedline}{\raisebox{2pt}{\tikz{\draw[-,black,dotted,line width = 0.75pt](0.,0mm) -- (5mm,0mm)}}}
\newcommand{\blackdashdottedline}{\raisebox{2pt}{\tikz{\draw[-,black,dashdotted,line width = 0.75pt](0.,0mm) -- (5mm,0mm)}}}
\newcommand{\Grayline}{\raisebox{2pt}{\tikz{\draw[-,Gray,solid,line width = 0.75pt](0.,0mm) -- (5mm,0mm)}}}
\newcommand{\Graydashedline}{\raisebox{2pt}{\tikz{\draw[-,Gray,dashed,line width = 0.75pt](0.,0mm) -- (5mm,0mm)}}}
\newcommand{\Graydashdottedline}{\raisebox{2pt}{\tikz{\draw[-,Gray,dashdotted,line width = 0.75pt](0.,0mm) -- (5mm,0mm)}}}
\newcommand{\Graydottedline}{\raisebox{2pt}{\tikz{\draw[-,Gray,dotted,line width = 0.75pt](0.,0mm) -- (5mm,0mm)}}}
\newcommand{\redline}{\raisebox{2pt}{\tikz{\draw[-,red,solid,line width = 0.75pt](0.,0mm) -- (5mm,0mm)}}}

\newcommand{\Limegreenline}{\raisebox{2pt}{\tikz{\draw[-,ForestGreen,solid,line width = 0.75pt](0.,0mm) -- (5mm,0mm)}}}

\newcommand{\blueline}{\raisebox{2pt}{\tikz{\draw[-,blue,solid,line width = 0.75pt](0.,0mm) -- (5mm,0mm)}}}

\newcommand{\reddashline}{\raisebox{2pt}{\tikz{\draw[-,red,dashed,line width = 0.75pt](0.,0mm) -- (5mm,0mm)}}}

\newcommand{\hollowcircle}{\raisebox{0pt}{\tikz{\draw[black,solid,line width = 0.5pt](2.5mm,0) circle [radius=0.75mm];}}}

\newcommand{\hollowbluecircle}{\raisebox{0pt}{\tikz{\draw[blue,solid,line width = 0.5pt](2.5mm,0) circle [radius=0.75mm]}}}
\newcommand{\hollowredcircle}{\raisebox{0pt}{\tikz{\draw[red,solid,line width = 0.5pt](2.5mm,0) circle [radius=0.75mm]}}}
\newcommand{\hollowcirclesolidline}{\raisebox{0pt}{\tikz{\draw[black,solid,line width = 0.5pt](2.5mm,0) circle [radius=0.75mm];\draw[-,black,solid,line width = 0.5pt](0.,0mm) -- (5mm,0mm)}}}

\newcommand{\hollowsquaresolidline}{\raisebox{0pt}{\tikz{\draw[black,solid,line width = 0.5pt](2.mm,0) rectangle (3.5mm,1.5mm);\draw[-,black,solid,line width = 0.5pt](0.,0.8mm) -- (5.5mm,0.8mm)}}}
\newcommand{\redsquare}{\raisebox{0pt}{\tikz{\draw[red,solid,line width = 0.5pt](2.mm,0) rectangle (3.5mm,1.5mm)}}}
\newcommand{\lightbluesquare}{\raisebox{0pt}{\tikz{\draw[teal,solid,line width = 0.5pt](2.mm,0) rectangle (3.5mm,1.5mm)}}}
\newcommand{\hollowsquareblack}{\raisebox{0pt}{\tikz{\draw[black,solid,line width = 0.5pt](2.mm,0) rectangle (3.5mm,1.5mm)}}}
\newcommand{\hollowsquarered}{\raisebox{0pt}{\tikz{\draw[red,solid,line width = 0.5pt](2.mm,0) rectangle (3.5mm,1.5mm)}}}
\newcommand{\lightgreensquare}{\raisebox{0pt}{\tikz{\draw[olive,solid,line width = 0.5pt](2.mm,0) rectangle (3.5mm,1.5mm)}}}
\newcommand{\hollowtrainglesolidline}{\raisebox{0pt}{\tikz{\draw[black,solid,line width = 0.5pt](2.mm,0) -- (3.5mm,0) -- (2.75mm,1.5mm) -- (2.mm,0);\draw[-,black,solid,line width = 0.5pt](0.,0.8mm) -- (5.5mm,0.8mm)}}}
\newcommand{\hollowtraingleblack}{\raisebox{0pt}{\tikz{\draw[black,solid,line width = 0.5pt](2.mm,0) -- (3.5mm,0) -- (2.75mm,1.5mm) -- (2.mm,0)}}}
\newcommand{\hollowtrainglered}{\raisebox{0pt}{\tikz{\draw[red,solid,line width = 0.5pt](2.mm,0) -- (3.5mm,0) -- (2.75mm,1.5mm) -- (2.mm,0)}}}
\newcommand{\greystar}{\raisebox{0pt}{\tikz{\draw[Gray,solid,line width = 0.5pt](0.75mm,1.5mm) -- (0.75mm,0);\draw[Gray,solid,line width = 0.5pt](0,0.75mm) -- (1.5mm,0.75mm)}}}
\newcommand{\blackstar}{\raisebox{0pt}{\tikz{\draw[black,solid,line width = 0.5pt](2.mm,0.8mm) -- (3.5mm,0.8mm); \draw[black,solid,line width = 0.5pt](2.75mm,1.5mm) -- (2.75mm,0);\draw[black,solid,line width = 0.5pt](2.325mm,1.25mm) -- (3.25mm,0.25mm);\draw[black,solid,line width = 0.5pt](2.325mm,0.25mm) -- (3.25mm,1.25mm)}}}
\newcommand{\hollowstarsolidline}{\raisebox{0pt}{\tikz{\draw[black,solid,line width = 0.5pt](2.mm,0.8mm) -- (3.5mm,0.8mm); \draw[black,solid,line width = 0.5pt](2.75mm,1.5mm) -- (2.75mm,0);\draw[black,solid,line width = 0.5pt](2.325mm,1.25mm) -- (3.25mm,0.25mm);\draw[black,solid,line width = 0.5pt](2.325mm,0.25mm) -- (3.25mm,1.25mm);\draw[-,black,solid,line width = 0.5pt](0.,0.8mm) -- (5.5mm,0.8mm)}}}

\newcommand{\tagarray}{\mbox{}\refstepcounter{equation}$(\theequation)$}

\shorttitle{Slip effect on lift and drag}
\shortauthor{N. I. K. Ekanayake \textit{et} al.}

\title{Lift and drag forces acting on a particle moving in the presence of slip and shear near a wall.}

\author{Nilanka. I. K. Ekanayake,
  Joseph D. Berry \and Dalton J. E. Harvie\corresp{\email{daltonh@unimelb.edu.au}}}

\affiliation{
Department of Chemical Engineering, The University of Melbourne, Victoria 3010 Australia
}

\begin{document}

\maketitle
\begin{abstract}
The lift and drag forces acting on a small spherical particle moving with a finite slip in single-wall-bounded flows are investigated via direct numerical simulations. The effect of slip velocity on the particle force is analysed as a function of separation distance for low slip and shear Reynolds numbers ($10^{-3} \leq Re_{\gamma}, Re_{\text{slip}} \leq 10^{-1}$) in both quiescent and linear shear flows. A generalised lift model valid for arbitrary particle-wall separation distances and $Re_{\gamma}, Re_{\text{slip}} \leq 10^{-1}$ is developed based on the results of the simulations. The proposed model can now predict the lift forces in linear shear flows in the presence or absence of slip,and in quiescent flows when slip is present. Existing drag models are also compared with numerical results for both quiescent and linear shear flows to determine which models capture near wall slip velocities most accurately for low particle Reynolds numbers. Finally, we compare the results of the proposed lift model to previous experimental results of buoyant particles and to numerical results of neutrally-buoyant (force-free) particles moving near a wall in quiescent and linear shear flows. The generalised lift model presented can be used to predict the behaviour of particle suspensions in biological and industrial flows where the particle Reynolds numbers based on slip and shear are $\mathcal{O}(10^{-1})$ and below.

\end{abstract}


\section{Introduction}
Small particles moving near a wall experience lift forces in a direction normal to the wall. In sheared flows these forces cause particles to migrate across fluid streamlines and cluster at different equilibrium locations away from the wall \citep{Segre62}. This passive particle migration, induced purely by hydrodynamic forces, is observed in biological flows causing, for example, cell migration in microvascular networks \citep{Leiderman}. This migration mechanism has also been exploited in the design of micro-scale cell sorting microfluidics \citep{DiCarlo09}, macro-scale particle deposition systems and shear enhanced membrane filtration devices \citep{vanderSman}. Accurate quantification of the lift forces acting on small particles is hence key in predicting particle distributions in both biological and non-biological suspension flows.

In this study, we are particularly interested in the lift forces acting on rigid spherical particles that are moving with a finite slip at low particle Reynolds numbers. In dilute systems, particle slip velocities can originate from a variety of forces, including fluid drag or buoyancy. These forces are often much higher in magnitude than the lift forces. For example, freely translating neutrally-buoyant particles experience a finite but relatively small slip velocity due to the wall-shear fluid drag force \citep{EkanayakeJFM1}. In contrast, buoyant particles sedimenting in vertical flows can experience much larger slip velocities due to strong buoyancy forces, which are further affected by wall-bounded fluid drag forces when particles are moving in close proximity to a wall. In both cases, the lift force acting on a particle strongly depends on the slip velocity, fluid shear rate and distance to the wall. A significant amount of theoretical work has examined lift forces for rigid particles at finite slip, however, a generalised wall-bounded correlation applicable for all particle-wall separation distances is not available.

The main objective of this work is to extend the existing slip-shear-wall based theoretical results given for $Re_\text{slip}, Re_\gamma \ll 1$ to larger particle Reynolds numbers up to $\mathcal{O}(10^{-1})$, directly relevant to particulate flows within small channels (i.e., particle migration in microfluidic devices). We use well resolved numerical simulations to define a general lift model for a particle experiencing slip in a shear flow for arbitrary particle-wall separation distances. For this, rigid spherical particles moving with finite slip tangential to a flat wall in quiescent and linear shear flows are considered for slip and shear Reynolds numbers in the range of $10^{-3}$ to $10^{-1}$. We consider both non-rotating and freely-rotating particles.

We first discuss the available slip based lift and drag models and their associated limitations, in \S\ref{sec:jfm2:LiftForce} and \S\ref{sec:jfm2:DragForce} respectively. Then, we define the numerical setup in \S\ref{sec:jfm2:NumericalSimulations}. In \S\ref{sec:jfm2:Results and Discussion}, we express our numerical results, for both quiescent and linear shear flows, as new lift correlations valid for arbitrary wall-particle separation distance. In \S \ref{sec:jfm2:Application}, we compare the results of these new lift correlations together with selected drag correlations against previous near wall experimental results for buoyant particles in quiescent and linear shear flows, as well as previous numerical results for force-free particles in linear shear flows.

\section{Existing Theories}\label{sec:jfm2:Theories}

In this section we outline the previous work related to lift and drag forces acting on rigid particles, and establish the limitations to be addressed in this study. For clarity, the force models are classified as unbounded (ub), wall-bounded outer-region (wb,out) and wall-bounded inner-region (wb,in) considering the wall and particle separation distance. Wall-bounded inner-region-based models consider a particle close enough to a wall such that the viscous effects are more significant than the inertial effects. Wall-bounded outer-region-based models consider a particle located far away from a wall, where both viscous and inertial effects are significant. The corresponding notation (ub, wb,out and wb,in) will appear in the superscript of each force coefficient. 

\subsection{Lift force}\label{sec:jfm2:LiftForce}

\subsubsection{Unbounded models}
The hydrodynamic lift force is an inertia-induced force that reduces to zero for rigid particles in Stokes flow \citep{Bretherton}. When inertia is present, a particle that either leads or lags the fluid flow can experience a lift force in unbounded linear shear flows. Accounting for this, \citet{Saffman} proposed an asymptotic expression for the lift force ($F_\text{L}$):
\begin{equation}
    {F_\text{L}^{\ast}} = \frac{F_{\text{L}} \rho}{\mu^2} = -\text{sgn}({\gamma^{\ast}})2.255\times\frac{9}{\pi} { {Re_\gamma}}^{\frac{1}{2}}Re_\text{slip} + \text{sgn}({\gamma^{\ast}})\frac{11}{8}Re_\gamma Re_\text{slip} - \pi Re_\omega Re_\text{slip}
    \label{eq:jfm2:saffmanlift}
\end{equation}
valid for low slip, shear and rotational Reynolds numbers ($Re_\text{slip}, Re_\gamma, Re_\omega \ll 1$). Here, 
\begin{equation*}
    Re_{\text{slip}} = \frac{|u_{\text{slip}}|a}{\nu},~~~~~~
    Re_{\gamma} = \frac{|\gamma| a^2}{\nu},~~~~~~Re_{\omega} = \frac{\omega a^2}{\nu},
\end{equation*}
and $u_{\text{slip}}$, $\omega$, $a$, $\nu$ and $\gamma$ are the particle slip velocity, particle angular rotation, particle radius, fluid kinematic viscosity and fluid shear rate, respectively. The shear rate normalised by the slip velocity $\gamma^{\ast} =\gamma a/u_\text{slip}$, depends on both the slip velocity and shear rate. The direction of the lift force is determined by the sign of $\gamma^{\ast}$. Saffman's model is an outer-region-based lift model, in which the boundary of the inner and outer-region is located at min$(L_{\text{G}},L_{\text{S}})$ from the particle. Here $L_{\text{S}}={\nu}/{|u_\text{slip}|}$ and $L_{\text{G}} = \sqrt{{\nu}/{|\gamma|}}$ are the Stokes and Saffman length scales respectively. In addition to the small particle Reynolds number constraints, inertial effects due to shear must be higher than the inertial effects generated by the slip velocity ($ \epsilon = \sqrt{\lvert{Re\gamma}\rvert}/Re_\text{slip} \gg 1$ or equivalently ${L_\text{G}} \ll {L_\text{S}}$) for the model to be valid. 

The first and second terms on the right hand side of the Eq. (\ref{eq:jfm2:saffmanlift}) are both due to fluid slip-shear effects, while the third term, similar to the lift model of \citet{Rubinow}, is due to the particle rotation. \citet{Saffman} illustrated that the lift force due to the rotation is less than that due to the shear by an order of magnitude, unless the rotational speed of a particle is much greater than the shear rate. Since the self induced rotation of a freely-rotating particle was shown to be small compared to the slip-shear lift for the condition $Re_\gamma \ll 1$, many studies neglect the third term when considering a freely translating and rotating particle. Additionally, the second term in Eq. (\ref{eq:jfm2:saffmanlift}) is less important than the first when $\epsilon \gg 1$ \citep{McLaughlin1}, and as a consequence, many outer-region studies focus solely on the first term of Eq. (\ref{eq:jfm2:saffmanlift}). Such models are referred to as first order models. 

\citet{Saffman}'s first order lift solution predicts a lift force in the direction of increasing fluid velocity for a lagging particle in a positive shear ($u_\text{slip} < 0, \gamma > 0$) or for a leading particle in a negative shear ($u_\text{slip} > 0, \gamma < 0$). Hereafter for convenience, any lift force acting in the direction of increasing fluid velocity in unbounded flows, will be defined as a positive lift. If both the slip and shear have the same sign (i.e., $\gamma^{\ast} > 0$), the lift direction reverses and Eq. (\ref{eq:jfm2:saffmanlift}) predicts a force in the direction of decreasing fluid velocity, which is a negative lift force.

\begin{figure}
\centering
\includegraphics[width=0.8\columnwidth]{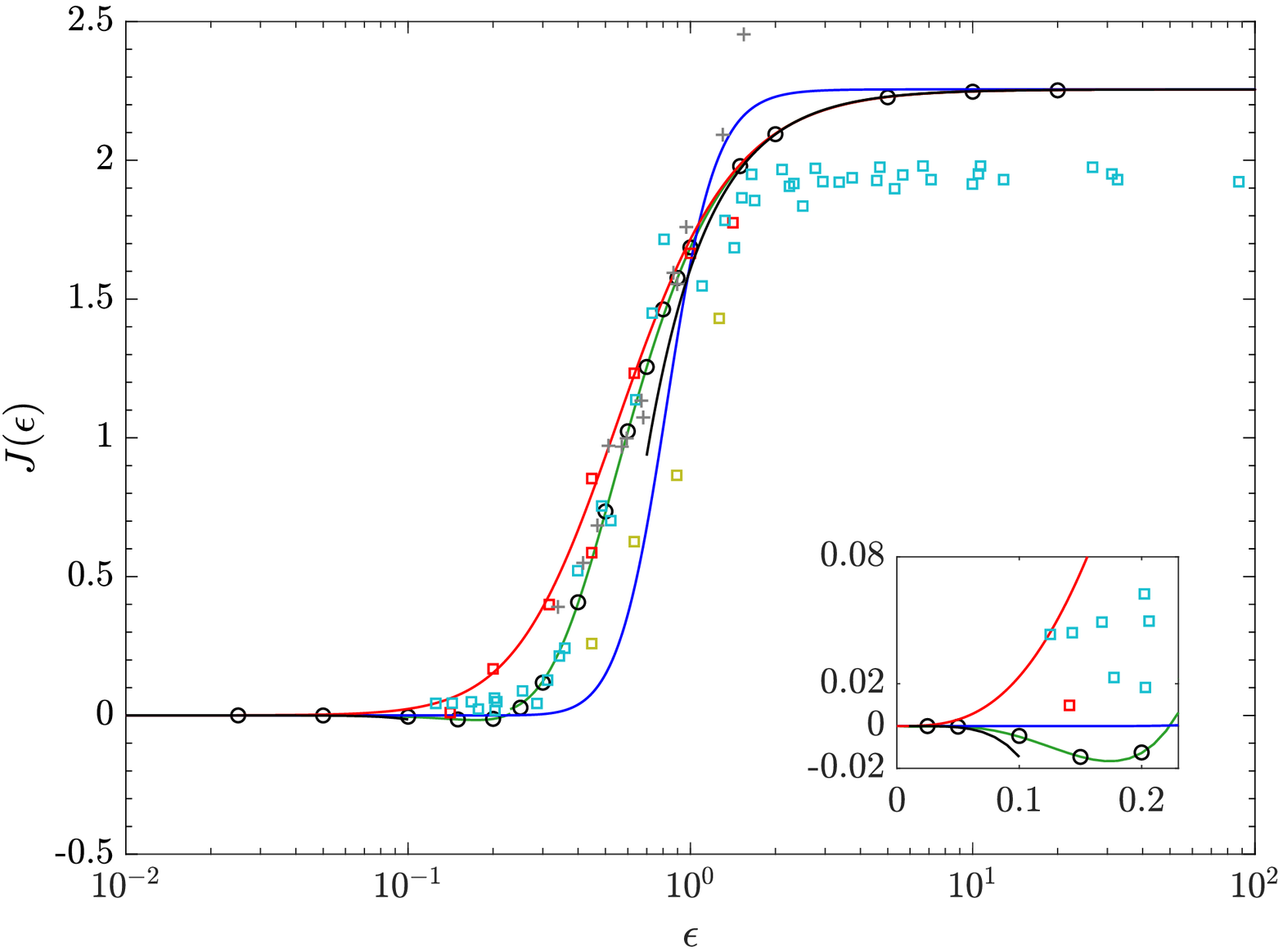}
\caption{Comparison of $J(\epsilon)$ values from experimental and DNS data for $Re_\text{slip} <1$ with empirical and theoretical correlations. Asymptotic solution (\protect\hollowcircle) and asymptotic limits (\protect\blackline) by \citet{McLaughlin1}. DNS: \citet{Legendre98} (\protect\redsquare), \citet{CherukatDandy} (\protect\lightbluesquare), \citet{Kurose99} (\protect\lightgreensquare). Experiments: \citet{CherukatGraham} (\protect\greystar), Empirical fittings by \citet{Mei} (\protect\blueline), \citet{Shi1} (\protect\Limegreenline) and \citet{Legendre98} (\protect\redline).} 
\label{fig:jfm2:Ju}
\end{figure}

Relaxing the constrains of $\epsilon \gg 1$ in \citet{Saffman}'s first order solution and using the Oseen approximation, \citet{McLaughlin1} and \citet{Asmolov90} independently proposed modified unbounded lift models in the form of,
\begin{align}
    {F_\text{L}^{\ast}} = -\text{sgn}({\gamma}^{\ast})\frac{9}{\pi} Re_\gamma^{\frac{1}{2}}Re_\text{slip} J(\epsilon)\label{eq:jfm2:mclaughlin1}\\
    \intertext{valid for non-rotating particles at $Re_\text{slip}, Re_\gamma \ll 1$. This force can be written in terms of a slip-shear lift force coefficient for unbounded flow, defined by} 
    {C}^\text{ub}_\text{L,2} = \frac{F_\text{L}^{\ast}}{-\text{sgn}(\gamma^{\ast})Re_\text{slip}^2} = \frac{9}{\pi} \epsilon J(\epsilon)\label{eq:unbounded_lift}
\end{align}
The function $J(\epsilon)$ needs to be evaluated analytically or numerically. Currently available expressions based on asymptotic solutions and empirical fitting functions for $J(\epsilon)$ are presented in table \ref{tabel:jfm2:Ju correlations}. These expressions, along with previous direct numerical simulation and experimental results, are plotted in figure \ref{fig:jfm2:Ju}. At the limit $\epsilon \rightarrow \infty$ (or equivalently at the limit $Re_\gamma \gg Re_\text{slip}$), $J(\epsilon)$ reduces to the \citeauthor{Saffman}'s limit of 2.255, whereas $J(\epsilon)$ decreases to zero rapidly as $\epsilon$ decreases.

In the \citet{McLaughlin1} and \citet{Asmolov90} studies the integral expression for $J(\epsilon)$ was evaluated numerically. In addition, \citet{McLaughlin1} also provided two analytical solutions for $J(\epsilon)$ at the limits of $\epsilon \ll 1$ and $\epsilon \gg 1$ (see figure \ref{fig:jfm2:Ju}). The values obtained from numerical evaluations suggested a positive $J(\epsilon)$ for $\epsilon > 0.23$ and a negative $J(\epsilon)$ for $\epsilon < 0.23$. Based on \citet{McLaughlin1}'s theoretical results, specifically given for the range $0.1<\epsilon<20$, \citet{Mei} proposed a fitting function for $J(\epsilon)$. In a similar vein, the same integral expression of $J(\epsilon)$ was re-evaluated numerically by \citet{Shi1} who proposed another fitting function to capture both negative and positive values accurately. Note however, these asymptotic solutions derived for $Re_\gamma, Re_\text{slip} \ll 1$ may not be valid for larger Reynolds numbers, specifically for the $\mathcal{O}(10^{-1})$ ranges relevant to this study.

Direct numerical simulation (DNS) studies, whereby the flow around an individual particle is simulated, provide better estimations of $J(\epsilon)$ for $Re_\text{slip}, Re_\gamma \lesssim 1$ in unbounded-linear shear flows, as they do not rely on the Oseen approximation \citep{Dandy, Legendre98,Kurose99,CherukatDandy}. \citet{Dandy} performed the first DNS study of the flow around a rigid sphere in an unbounded linear shear flow. However, the values obtained for the lift at small Reynolds numbers were later shown by subsequent DNS studies to be significantly in error due to the small domain size (25 particle radii) employed \citep{CherukatDandy,Legendre98}. \citet{Legendre98} performed simulations for a clean spherical bubble using large domain sizes (100 particle radii and 200 particle radii \citep{Shi1}). The numerical data and theoretical estimations from \citet{McLaughlin1} were in good agreement for $\epsilon \geq 0.5$, however, the negative $J(\epsilon)$ values for $0<\epsilon<0.23$ predicted by the theoretical studies were not observed. Citing reasons for this discrepancy, \citet{Legendre98}, and later \citet{TakemuraMagnaudet2}, explained that the theoretical integral expression obtained for $J(\epsilon)$ is based on the Oseen approximation, which is not sufficiently accurate to evaluate the small lift forces that exists at low shear rates. They illustrated that this approximation cannot capture the higher order terms in the lift force expansion (i.e., second term in Eq. (\ref{eq:jfm2:saffmanlift})) which is important when $\epsilon$ is very small ($Re_\text{slip} \gg \sqrt{Re_\gamma}$) \citep{McLaughlin1,Legendre98}. Comparing their numerical data with theoretical values, \citet{Legendre98} suggested that the lower bound of validity in the asymptotic solution is $\epsilon \approx 0.7$. In the same study an empirical fitting for $J(\epsilon)$ was suggested, based on their numerical results for $0.2 < \epsilon < 0.6$ at $Re_\text{slip} < 1$ and \citet{McLaughlin1}'s theoretical results for $\epsilon > 0.8$. The resulting correlation predicts a positive lift force for all $\epsilon$ values. Both \citet{Kurose99} and \citet{CherukatDandy} performed DNS simulations specifically for a rigid sphere translating in unbounded linear shear flows. \citet{CherukatDandy} used large domains (75 and 105 particle radii) and tested for low slip and shear Reynolds number combinations ($0.01 < Re_\text{slip} < 1$, $0.01<Re_\gamma<0.025$). The computed $J(\epsilon)$ values were positive for all $\epsilon$ at $Re_\text{slip}, Re_\gamma < 1$ as illustrated in figure \ref{fig:jfm2:Ju}. However, the results for $\epsilon \geq 2$ showed deviations from other asymptotic predictions, and this discrepancy is explained as a domain truncation error \citep{CherukatDandy}. \citet{Kurose99} performed simulations for relatively large slip Reynolds numbers $0.25< Re_\text{slip} < 250$ and hence employed relatively small domain sizes (10, 20 particle radii), and the computed $J(\epsilon)$ values agreed well with other numerical studies. 

\renewcommand{\arraystretch}{1.2}
\begin{table}
    \centering{
    \resizebox{\linewidth}{!}{%
    \begin{tabular}{lllc} 
    Study & $J(\epsilon)$ & Limits & Eq.\\
    \hline
    \multirow{2}{*}{\citet{McLaughlin1}} & $-32\pi^{2}\epsilon^{5}ln(1/\epsilon^{2})$ & $\epsilon \ll 1$ & \multirow{2}{*}\quad\tagarray\label{McL}\\
    & $2.255-0.6463\epsilon^{-2}$ & $\epsilon \gg 1$\\\\
    \citet{Mei} & $0.6765\{1+\text{tanh}[2.5(lg\epsilon+0.191)]\}\{0.667+\text{tanh}[6(\epsilon-0.32]\}$ & $0.2 < \epsilon < 20$ & \quad\tagarray\label{Mei}\\\\
    \citet{Legendre98} & $2.255(1+0.20\epsilon^{-2})^{-3/2}$ & $0 < \epsilon < 10$ & \quad\tagarray\label{Leg}\\\\
    \multirow{2}{*}{\citet{Shi1}} & $-0.04\epsilon + 2.05\epsilon^2-32.2\epsilon^3+106.8\epsilon^4$ & $\epsilon \leq 0.23$\\
    & $2.255(1+0.02304\epsilon^{-2})^{-12.77}$ & $\epsilon > 0.23$ & \multirow{2}{*}\quad\tagarray\label{Shi}\\
    \end{tabular}}
    }
    \caption{Correlations for $J(\epsilon)$}
  \label{tabel:jfm2:Ju correlations}
\end{table}

Experimentally, \citet{CherukatGraham} investigated the variation of $J(\epsilon)$ at small $Re_\gamma$ and $Re_\text{slip}$ numbers in unbounded flows. The migration velocities of a small negatively buoyant particle sedimenting in a linear shear flow were measured. To allow comparison with theory, we have converted these migration velocities to $J(\epsilon)$ using Eq.(\ref{eq:jfm2:mclaughlin1}) and Stokes Law, and plotted these results in figure \ref{fig:jfm2:Ju}.  As illustrated in the figure, the experimental values for $J(\epsilon)$ closely follow the asymptotic theories up to $\epsilon \sim 1$, but beyond this there is a difference between the two results. \citet{CherukatGraham} explained that the inconsistencies between experimental and theoretical values in this region are due to experimental errors in the measurements of low migration velocities at low $Re_\text{slip}$. Consistent with the DNS results, negative $J(\epsilon)$ were not observed in any of these experiments, specifically for $\epsilon <0.23$. This again suggests that any analytical solution or empirical correlation based on Oseen's approximation are invalid for low $\epsilon$.


\subsubsection{Wall-bounded models}
In bounded flows (i.e., near a wall), particles moving at a finite slip velocity experience an additional lift force even in the absence of shear. This wall-slip lift is greatest near the wall and reduces rapidly to zero away from the wall. Note that while a particle slip velocity can be in any direction relative to a wall, in this study we only consider slip-lift due to slip velocity in the direction parallel to a wall.

\paragraph{Outer-region}

The wall-slip lift for a spherical particle sedimenting in a quiescent fluid ($Re_\gamma = 0$) with a single wall located in the outer-region ($l>L_\text{S}$) was first investigated by \citet{VasseurCox2}. Here $l$ is the distance between the particle centre and the wall. Singular perturbation techniques were used to determine the migration velocity and the equivalent lift force was then calculated using Stokes law. The deduced lift force valid for $Re_\text{slip} \ll 1$ was given as:
\begin{equation}
    {F_\text{L}^{\ast}} = C_\text{L,3}^\text{wb,out}(l/L_\text{S})Re_\text{slip}^2
    \label{eq:sliplift-outer}
\end{equation}
The integral expression for $C_\text{L,3}^\text{wb,out}$ was evaluated numerically as a function of separation distance normalized by Stokes length $l/L_\text{S}$. In the same study, the asymptotic behaviour at small and large values of ${L}_\text{S}$ was obtained analytically considering the inner and outer boundary limits of the outer-region ($l/L_\text{S} \ll 1$ and $l/L_\text{S} \gg 1$ respectively). Although a rotating sphere was originally considered, \citet{VasseurCox2} illustrated that the calculated lift force is independent of rotation as long as the angular velocity is less than $\mathcal{O}(Re_\text{slip})$ in the outer-region. Hence Eq. (\ref{eq:sliplift-outer}) is applicable for a non-rotating sphere as well. Several studies have developed empirical fitting correlations for $C_\text{L,3}^\text{wb,out}$ by solving the integral expression for $C_\text{L,3}^\text{wb,out}$ numerically \citep{Takemura2003,Takemura,Shi2}. These expressions are listed in table \ref{table:cl2wbouter}, in addition to the analytical solutions obtained for the asymptotic limits, and the predictions of these correlations show that a leading or lagging particle in quiescent flows always moves away from the wall. Hence the deduced lift coefficient $C_\text{L,3}^\text{wb,out}$ is positive irrespective of the slip velocity direction, but reduces to zero as $l/L_\text{S} \rightarrow \infty$.

\citet{VasseurCox2} and later \citet{Takemura} conducted experiments to measure the migration velocity of a rigid particle sedimenting in a quiescent flow. While the first study obtained migration velocities of a particle falling relatively far away from the wall, the latter study focused mainly on obtaining experimental results for the inner-region. The experimental measurements of \citet{VasseurCox2} obtained mainly for $l/L_\text{S} > 1$ agreed well with the outer-region-based $C_\text{L,3}^\text{wb,out}$ correlations.

\renewcommand{\arraystretch}{1.2}
\begin{table}
    \centering{
    \resizebox{\linewidth}{!}{%
    \begin{tabular}{llll} 
    Study & $C_\text{L,3}^\text{wb,out}$ & Limits & Eq.\\
    \hline
    \multirow{2}{*}{\citet{VasseurCox2}} & $({9\pi}/{16})\big[1-{11}/{32}({l}/{L_\text{S}})\big]$ & ${l}/{L_\text{S}} \ll 1$ & \multirow{2}{*}\quad\tagarray\label{VC2}\\
    & $({9\pi}/{4})\big[({l}/{L_\text{S}})^{-2} + 2.21901({l}/{L_\text{S}})^{-5/2}\big]$ & ${l}/{L_\text{S}} \gg 1$\\\\
    \multirow{2}{*}{\citet{Takemura2003}} & $\big[{9\pi}/{16} + 2.89\pi\times 10^{-6}({l}/{L_\text{S}})^{4.58}\big]e^{-0.292({l}/{L_\text{S}})}$ & $ 0 < {l}/{L_\text{S}} < 10$ & \multirow{2}{*}\quad\tagarray\label{TM2003}\\
    & $4.47\pi({l}/{L_\text{S}})^{-2.09}$ & $ 10 \leq {l}/{L_\text{S}} < 100$\\\\
    {\citet{Takemura}} & ${18\pi}\big[32+2\big({l}/{L_\text{S}}\big)+3.8({l}/{L_\text{S}})^2+0.049({l}/{L_\text{S}})^3]^{-1}$ & $0<l/L_\text{S} < 10$ & \quad\tagarray\label{TM2004}\\\\
    \multirow{2}{*}{\citet{Shi2}} & $({9\pi}/{16})\big[1+0.13({l}/{L_\text{S}})\big(({l}/{L_\text{S}})+0.53\big)\big]^{-1}$ & $ 0 < {l}/{L_\text{S}} < 10$ & \multirow{2}{*}\quad\tagarray\label{ShiCL3}\\
    & $4.47\pi({l}/{L_\text{S}})^{-2.09}$ & $ 10 \leq {l}/{L_\text{S}} < 100$
    \end{tabular}}
    }
    \caption{Correlations for outer-region-based wall-slip lift coefficient $C_\text{L,3}^\text{wb,out}$}
  \label{table:cl2wbouter}
\end{table}

The presence of a wall also affects the slip-shear lift force acting on a particle in a linear shear flow. A non-rotating sphere in a single wall-bounded linear shear flow with the wall lying in the outer-region was first investigated by \citet{Drew88}, and later by \citet{Asmolov89} and \citet{McLaughlin2}. The latter two studies used the method of matched asymptotic expansions, and considered a leading and a lagging particle in a positive shear field. These cases correspond to ${\gamma}^{\ast}>0$ and ${\gamma}^{\ast}<0$ respectively. In the outer-region, the effect of rotation was shown to be less significant,hence the developed models are applicable for both freely-rotating or non-rotating particles. \citet{Drew88} considered the problem in the limit of $\epsilon \gg 1$ while \citet{McLaughlin2} and \citet{ Asmolov89} considered a range of $\epsilon$ values. Based on the Oseen approximation, an analytical solution for $l \gg L_{\text{G}}$ and $\epsilon \gg 1$ (but not necessarily $Re_\text{slip} = 0$) was also provided in the \citet{McLaughlin2} study. 

For $Re_\gamma, Re_\text{slip} \ll 1$ and $l \gg$ min$(L_{\text{G}},L_{\text{S}})$ (i.e., in the outer-region) the wall-bounded slip-lift force can be presented as,
\begin{align}
    {F_\text{L}^{\ast}} &=- \text{sgn}({\gamma}^{\ast})C_\text{L,2}^\text{wb,out}Re_\text{slip}^2
    \label{eq:slipshearlift-outer}
\end{align}
The numerical values obtained for $C_\text{L,2}^\text{wb,out}$ by solving Airy functions indicated that the unbounded slip-shear lift varies as $l/L_\text{G}$ changes \citep{McLaughlin2,Asmolov89}. For $\epsilon \gg 1$, $C_\text{L,2}^\text{wb,out}$ monotonically reduced from the unbounded values to near zero values for small enough $l/L_\text{G}$,  irrespective of the sign of ${\gamma}^{\ast}$. \citet{McLaughlin2} showed that these near zero values are similar to the outer boundary values of the inner-region solutions of \citet{CoxHsu} when $l/L_\text{G} <1$ and $\epsilon > 1$. Based on the numerical data tabulated in \citet{McLaughlin2}, for both ${\gamma}^{\ast} > 0$ and ${\gamma}^{\ast} < 0$, and considering the asymptotic inner-region solution of \citet{CoxHsu} valid for $l^{\ast} < L_\text{G}^{\ast}$, \cite{TakemuraMagnaudet1} and \citet{Shi2} both proposed semi-empirical fits for $C_\text{L,2}^\text{wb,out}$ for $\epsilon > 1$ in the form of:
\begin{align}
    C_\text{L,2}^\text{wb,out} = f(\epsilon, l/L_\text{G})C_\text{L,2}^\text{ub}
    \label{eq:slipshearlift-f}
\end{align}
Table \ref{table:CL2,Ub_correlations} summarises the available theoretical and empirical correlations for $f(\epsilon, l/L_\text{G})$.

Noting that when shear is negligibly small ($\epsilon \rightarrow 0$, or $Re_\gamma \rightarrow  0$) the lift contribution should be entirely due to the disturbance produced by the wall and slip effects, several studies have attempted to combine Eq. (\ref{eq:slipshearlift-outer}) and Eq. (\ref{eq:sliplift-outer}) such that $\text{sgn}(\gamma^{\ast})C_\text{L,2}^\text{wb,out}$ tends towards to $C_\text{L,3}^\text{wb,out}$ in the limit of $\epsilon \rightarrow 0$. Using Eqs.(\ref{TM2004}) and (\ref{f_TM}) for $C_\text{L,3}^\text{wb,out}$ and $C_\text{L,2}^\text{wb,out}$ respectively, \citet{TakemuraMagnaudet1} combined these two coefficients using a fitting function ($f_{2}({\epsilon,l/L_\text{S}})$) as below:
\begin{subequations}
\begin{gather}
    F_\text{L}^{\ast} = \bigg(-\text{sgn}({\gamma}^{\ast})C_\text{L,2}^\text{wb,out} + f_2({\epsilon,l/L_\text{S}})C_\text{L,3}^\text{wb,out}\bigg)Re_{\text{slip}}^2\\
    \intertext{where}
    f_{2}({\epsilon,l/L_\text{S}}) = \text{exp}{(-0.22 \epsilon^{3.3} (l/L_\text{S})^{2.5})}\\
    \intertext{Here,}
    C_\text{L,23}^{\text{wb,out}} = \frac{F_\text{L}^{\ast}}{Re_{\text{slip}}^2} = -\text{sgn}({\gamma}^{\ast})C_\text{L,2}^\text{wb,out} + f_2({\epsilon,l/L_\text{S}})C_\text{L,3}^\text{wb,out}
\end{gather}
\label{eq:slipshearlift-outercl2cl3}%
\end{subequations}
is the corresponding force coefficient, with the lift normalised by the slip Reynolds number. This outer-region-based lift model given by Eq. (\ref{eq:slipshearlift-outercl2cl3}) performs well for small and intermediate $\epsilon \gtrsim 1$ \citep{TakemuraMagnaudet1,Shi2}, however, it is worth noting that this model predicts a zero lift force in the absence of slip which is not necessarily accurate \citep{EkanayakeJFM1}.

The lift force acting on a spherical particle translating with zero-slip velocity ($Re_\text{slip} =0$ or equivalently $\epsilon = \infty$) in the outer-region of a positive shear flow has also been studied theoretically and numerically. Asymptotic studies for $Re_\gamma \ll 1$ \citep{Asmolov99} and DNS studies for $10^{-3}< Re_\gamma < 10^{-1}$ \citep{EkanayakeJFM1} find a lift force that decays rapidly to zero with increasing separation distance. Based on numerical lift results, \citet{EkanayakeJFM1} proposed an outer-region-based lift model accounting for both shear and wall effects:
\begin{subequations}
\begin{gather}
    {F_\text{L}^{\ast}} = C_\text{L,1}^\text{wb,out}(l/L_\text{G})Re_\gamma^2
    \intertext{where}
    C_\text{L,1}^\text{wb,out} = 2.231 e^{(-0.1054(l/L_\text{G})^2-0.3859(l/L_\text{G}))}
    \intertext{for non-rotating particles and}
    C_\text{L,1}^\text{wb,out} = 1.982 e^{(-0.115(l/L_\text{G})^2-0.2771(l/L_\text{G}))}
\end{gather}
\label{eq:shearlift-outer}%
\end{subequations}
for freely-rotating particles, with $C_\text{L,1}^\text{wb,out}$ approaching zero as $l/L_\text{G}$ increases.  Similar to the slip based lift coefficient ($C_\text{L,3}^\text{wb,out}$), the shear based lift coefficient $C_\text{L,1}^\text{wb,out}$ remains positive for both negative and positive shear rates, favouring particle migration away from the wall.

\renewcommand{\arraystretch}{1.2}
\begin{table}
    \centering{
    \resizebox{\linewidth}{!}
    {%
    \begin{tabular}{llll} 
    Study & $f(\epsilon, l/L_\text{G})$ & Limits & Eq.\\
    \hline
    {\citet{McLaughlin2}} & $1-1.8778 (l/L_\text{G})^{-5/3}/J(\epsilon)$ & $l/L_\text{G}\gg1, \epsilon \gg 1$ & \quad\tagarray\label{f_ML}\\\\
    {\citet{CoxHsu}} & $11/96\pi^2(l/L_\text{G})/J(\epsilon)$ & $l/L_\text{G}\ll1$ & \quad\tagarray\label{f_Cox}\\\\
    {\citet{TakemuraMagnaudet1}} & $1-\text{exp}\Big[-\dfrac{11}{96}\pi^2(l/L_\text{G})/J(\epsilon)\Big]$ & $l/L_\text{G}>1, \epsilon > 1$ & \quad\tagarray\label{f_TM}\\\\
    \multirow{2}{*}{\citet{Shi2}} & $1-\text{exp}\Big[-\dfrac{11}{96}\pi^2\dfrac{70}{70+(l/L_\text{G})^{1.378}}(l/L_\text{G})/\lvert J(\epsilon) \rvert \Big]$ & ${l/L_\text{G}} < 15, \epsilon > 1$ & \quad\tagarray\label{f_Shi}\\
    & $1-1.8778L_\text{G}^{-5/3}/\lvert J(\epsilon)\rvert$ & $ {l}/{L_\text{G}} \geq 15, \epsilon > 1$\\
    \end{tabular}}
    }
    \caption{Numerical correlations for $f(\epsilon, L_\text{G})$}
  \label{table:CL2,Ub_correlations}
\end{table}

\paragraph{Inner-region}
Inner-region-based models require a particle to be located close to the wall such that $l \ll$ min$(L_{\text{G}},L_{\text{S}})$ and $Re_{\text{slip}}, Re_{\gamma} \ll 1$ \citep{CoxBrenner}. In these models the lift force on both freely-rotating and non-rotating particles is obtained by coupling the two flow disturbances that originate from particle slip and fluid shear in a non-linear manner. These inner-region-based lift models present the lift as \citep{CherukatMcLaughlin,Magnaudet1}:
\begin{equation}
     F_\text{L}^{\ast} = C_\text{L,1}^\text{wb,in}Re_{\gamma}^2+\text{sgn}(\gamma^\ast)C_\text{L,2}^\text{wb,in}Re_{\gamma}Re_{\text{slip}}+C_\text{L,3}^\text{wb,in}Re_{\text{slip}}^2
    \label{eq:liftinner}
\end{equation}
where the three lift coefficients, $C_\text{L,1}^\text{wb,in},C_\text{L,2}^\text{wb,in}$ and $C_\text{L,3}^\text{wb,in}$  are functions of separation distance. The first and last terms on the right hand side of Eq. (\ref{eq:liftinner}) originate from the disturbance induced by the presence of the wall in a shear flow field and by the presence of the wall in quiescent flow, respectively. The corresponding lift coefficients $C_\text{L,1}^\text{wb,in}$ and $C_\text{L,3}^\text{wb,in}$ are therefore associated with a force in the absence of slip ($Re_\text{slip}=0$) and a force in the absence of slip ($Re_\gamma=0$), respectively. The second term depends on both the slip velocity and shear rate, and the corresponding coefficient $C_\text{L,2}^{\text{wb,in}}$ is associated with a force when both the slip and shear are of the same order of magnitude. The first and last terms in Eq. (\ref{eq:liftinner}) produce forces directed away from the wall, resulting in a positive lift whereas the lift force due to the second term depends on both slip and shear rate directions, with the direction of this force captured by the sign of $(\gamma^*)$. The available correlations for $C_\text{L,2}^\text{wb,in}$ and $C_\text{L,3}^\text{wb,in}$ are summarised in table \ref{table:InnerLiftSummary}. Correlations available for $C_\text{L,1}^\text{wb,in}$ are tabulated in our previous study \citep{EkanayakeJFM1} and hence not detailed further here. 

In the theoretical context, \citet{CoxBrenner} were the first to obtain an implicit expression for the lift forces in the inner-region by using point force approximations at $l/a \gg 1$. Later \citet{CoxHsu} simplified this and presented closure expressions for lift coefficients with the leading order term proportional to ${l/a}$. The model is valid only when the separation distance is large compared to the sphere radius ($l/a \gg 1$). Accounting for the finite size of the particle, several other inner-region studies considered higher order contributions to the flow disturbances, and proposed lift correlations that are valid for a particle almost in contact with the wall ($l/a \gtrsim 1$) \citep{Leighton,Krishnan,CherukatMcLaughlin,Magnaudet1}. Unlike for the outer-region models, the effect of rotation is significant on lift coefficients within the inner-region models, particularly when the particle is close to the wall. Overall, as the inner-region models require a particle to be close to a wall, these models cannot be used to predict unbounded results as ${l/a}$ becomes large even at $Re_{\text{slip}}, Re_{\gamma} \ll 1$.

To summarise, the above analysis on existing lift force theories shows that all the presented lift models are limited to specific ranges of wall separation distance, fluid shear rate and particle slip velocity. For example, lift coefficients currently available for quiescent flows ($C_\text{L,3}$) are region specific (i.e., either inner-region or outer-region based) and do not account for any slip based inertial corrections particular when a particle translates closer to a wall (i.e., towards and within the inner region). For linear shear flows, existing slip-shear based lift coefficients ($C_\text{L,2}$) are also region specific and hence cannot capture the slip or shear based inertial dependence when a particle translates closer to a wall. The $C_\text{L,2}$  correlations that capture the inertial dependence of slip and shear are always limited to the systems where slip is stronger than shear, and thereby fail to capture the lift forces when shear is strong (i.e., freely translating neutrally-buoyant particles in shear flows). Hence, a generalised lift model valid for arbitrary particle-wall separation distances is necessary to make accurate predictions of particle distributions in industrial applications where  $Re_{\text{slip}},Re_\gamma<10^{-1}$. 

\begin{landscape}
\centering
\vspace*{\fill}
\begin{vtable}{{\centering}
\centering
\resizebox{\linewidth}{!}{%
    \centering
    \begin{tabular}{llllc}
    Study & $C_\text{L,2}^\text{wb,in}$ & $C_\text{L,3}^\text{wb,in}$ & Comments & Eq. \\
    \hline
    \\
    \multirow{4}{*}{\citet{CoxHsu}$^\dagger$} & $-\dfrac{66\pi}{64}\bigg[\bigg(\dfrac{l}{a}\bigg) + \dfrac{374\pi}{1056}\bigg]$  & $\dfrac{18\pi}{32}$ & Non-rotating, ${l/a} \gg 1$ & \multirow{2}{*}\quad\tagarray\label{Cox1}\\
    \\
    & $-\dfrac{66\pi}{64}\bigg[\bigg(\dfrac{l}{a}\bigg) + \dfrac{443\pi}{528}\bigg]$ & $\dfrac{18\pi}{32}$ & Freely-rotating, ${l/a} \gg 1$ & \multirow{2}{*}\quad\tagarray\label{Cox2}\\
    \\
    \hline
    \\
    {\citet{CherukatMcLaughlin}} & $-3.2397\bigg(\dfrac{l}{a}\bigg)-1.1450-2.0840\bigg(\dfrac{a}{l}\bigg)+0.9059\bigg(\dfrac{a}{l}\bigg)^{2}$ & $1.7716+0.2160\bigg(\dfrac{a}{l}\bigg)-0.7292\bigg(\dfrac{a}{l}\bigg)^{2}+0.4854\bigg(\dfrac{a}{l}\bigg)^{3}$ &Non-rotating, ${l/a} \gtrsim 1$ & \quad\tagarray\label{CM1}\\
    \\
    {\citet{CherukatMcLaughlinCorrection}} & $-3.2415\bigg(\dfrac{l}{a}\bigg)-2.6729-0.8373\bigg(\dfrac{a}{l}\bigg)+0.4683\bigg(\dfrac{a}{l}\bigg)^{2}$ & $1.7669+0.2885\bigg(\dfrac{a}{l}\bigg)-0.9025\bigg(\dfrac{a}{l}\bigg)^{2}+0.5076\bigg(\dfrac{a}{l}\bigg)^{3}$ & Freely-rotating, ${l/a} \gtrsim 1$ & \quad\tagarray\label{CM2}\\
    \\
    \hline
    \\
    \multirow{3}{*}{\citet{Krishnan}} & $-5.534$ & $1.755$ & Non-rotating, ${l/a}=1$ & \quad\tagarray\label{KL1}\\
    \\
    & $-2.091$ & $0.236$& Freely-rotating, ${l/a}=1$&\quad\tagarray\label{KL2}\\
    \\
    \hline
    \\
    {\citet{Magnaudet1}} & $\dfrac{-66\pi}{64}\bigg[\bigg(\dfrac{l}{a}\bigg)+\dfrac{443}{528}+\dfrac{52}{55}\bigg(\dfrac{a}{l}\bigg)\bigg]$ & $\dfrac{18\pi}{32}\bigg[1+0.1875\bigg(\dfrac{a}{l}\bigg)-0.168\bigg(\dfrac{a}{l}\bigg)^2\bigg]$ & Freely-rotating, ${l/a} \gtrsim 1$&\quad\tagarray\label{MD1}\\
    \\
\end{tabular}}
\caption{Slip based lift coefficients ($C_\text{L,2}^\text{wb,in}$ and $C_\text{L,3}^\text{wb,in}$) of inner-region studies.}
\label{table:InnerLiftSummary}}
\end{vtable}%
\footnotetext{$^\dagger$ minor corrections were provided by Lovalenti in the Appendix of \citet{CherukatMcLaughlin}}

\end{landscape}
\subsection{Drag force}\label{sec:jfm2:DragForce}
\subsubsection{Unbounded models}
The drag force acting on a rigid sphere translating with a finite slip velocity in an unbounded quiescent flow was first examined by \citet{Stokes851}. The study only considered the inner-region of the disturbed flow and assumed $Re_\text{slip} \ll 1$. The finite inertial effects in the outer-region of the disturbed flow were later analysed by \citet{Oseen1910}, who proposed a first order slip based inertial correction to the Stokes expression. Accounting for both inner and outer-regions of the disturbed flow, a higher order inertial correction for the drag force was suggested by \citet{Proudman57}, using a matched asymptotic method. The drag force predicted by \citeauthor{Proudman57}'s model reduces to Stokes' expression or Oseen's drag results, depending on the magnitude of $Re_\text{slip}$. Of note, these theoretical models are strictly limited to $Re_\text{slip} \ll 1$ and their predictions rapidly deviate from the measured drag forces for $Re_\text{slip} > 1$. Therefore, for $Re_\text{slip} \gtrsim 1$, empirical inertial corrections based on experimental and numerical data are more commonly used to capture the drag force variation \citep{Schiller33, Clift}. The drag force ($F_\text{D}$) acting on a spherical particle with a finite slip in an unbounded flow is generally presented as:
\begin{align}
    {{F}_\text{D}^\ast} = \frac{{F}_\text{D} \rho}{\mu} =  - \text{sgn} (u_\text{slip})Re_\text{slip}C_\text{D,2}^\text{ub}(Re_\text{slip})
    \label{eq:drag_ub}    
\end{align}
where $C_\text{D,2}^\text{ub}$ is the unbounded drag coefficient. Various theoretical and empirical correlations for $C_\text{D,2}^\text{ub}$, particularly for $Re_\text{slip} \lesssim 10$, are listed in the table \ref{unboundeddrag}.
\renewcommand{\arraystretch}{1.2}
\begin{table}
    \centering{
    \resizebox{\linewidth}{!}{%
    \begin{tabular}{llll} 
    Study & $C_\text{D,2}^\text{ub}(Re_\text{slip})$ & Limits & Eq. \\
    \hline
    \citet{Stokes851} & $6\pi$ & $Re_\text{slip} \ll 1 $ & \quad\tagarray\label{IC_Stokes}\\
    \citet{Oseen1910} & $6\pi (1+{3}/{8}Re_\text{slip})$ & $Re_\text{slip} < 1$ & \quad\tagarray\label{IC_Oseen}\\
    \citet{Proudman57} & $6\pi (1+{3}/{8}Re_\text{slip} + {9}/{40}Re_\text{slip}^{2}\text{ln} (Re_\text{slip}))$ & $Re_\text{slip} < 1$ & \quad\tagarray\label{IC_Proudman}\\
    \citet{Schiller33} & $6\pi \big(1+{1}/{6}Re_\text{slip}^{2/3}\big)$ & $Re_\text{slip} < 800$ & \quad\tagarray\label{IC_Schiller}\\
    \citet{Clift} & $6\pi(1+0.1315 Re_\text{slip}^{0.82-0.05\text{log}_{10}Re_\text{slip}})$ & $0.01 < Re_\text{slip} < 20$ & \quad\tagarray\label{IC_Clift}\\
    \end{tabular}}
    }
    \caption{Theoretical and empirical inertial corrections for $C_\text{D,2}^\text{ub}$}
  \label{unboundeddrag}
\end{table}

In unbounded linear shear flows, the effect of shear on the drag force is extremely weak for small slip Reynolds numbers ($Re_\text{slip}\lesssim 1$) \citep{Kurose99,Legendre98,Dandy}. However, for relatively large slip values ($Re_\text{slip} > 5$) and $Re_\gamma/Re_\text{slip} \sim \mathcal{O}(1)$, a noticeable effect from shear on the drag force occurs \citep{Kurose99}. For these large slip velocities theoretical arguments predict the drag as \citep{Legendre98}:
\begin{align}
    {{F}_\text{D}^\ast} = -\text{sgn} (u_\text{slip})Re_\text{slip}C_\text{D,2}^\text{ub}(Re_\text{slip})[1+K_{0}(Re_\gamma/Re_\text{slip})^2]
    \label{eq:drag_ub_inertial}    
\end{align}
The numerical results of \citet{Kurose99} suggested that $K_{0}$ is of order of unity for $Re_\text{slip}) \sim \mathcal{O}(1)$ and $K_{0} \simeq 0$ for $Re_\text{slip} \ll 1$. 


\subsubsection{Bounded models}
The effect of walls on the drag force was first examined by \citet{Faxen22} for a particle translating with a finite slip velocity parallel to a wall. The study considered a non-inertial ($Re_\text{slip} \ll 1$), quiescent flow ($Re_\gamma = 0$) with the walls located in the inner-region of the disturbed flow of the particle. In Faxen's study, the unbounded drag model, Eq. (\ref{eq:drag_ub}) was modified to incorporate wall effects via \citep{Happel}:
\begin{align}
    {{F}_\text{D}^\ast} = -\text{sgn} (u_\text{slip})Re_\text{slip}\big[C_\text{D,2}^\text{ub}(Re_\text{slip})+C_\text{D,2}^\text{wb,in}(a/l)\big]\label{eq:drag_wb_faxen}    \\
    \intertext{Thus, the net drag coefficient in a quiescent flow can be written as;}
    C_\text{D,2} = \frac{{F}_\text{D}^\ast}{ -\text{sgn} (u_\text{slip})Re_\text{slip}} = C_\text{D,2}^\text{ub}+C_\text{D,2}^\text{wb,in}\label{eq:drag_wb_faxen_cof}
\end{align}
The wall-bounded drag coefficient derived by Faxen, $C_\text{D,2}^\text{wb,in}$ consists of higher order terms of separation distance up to $\mathcal{O}((a/l)^5)$ in the drag force expansion.
\begin{align}
    \frac{C_\text{D,2}^\text{wb,in}}{6 \pi} = \Bigg[1-\frac{9}{16}\bigg(\frac{a}{l}\bigg)+\frac{1}{8}\bigg(\frac{a}{l}\bigg)^3-\frac{45}{256}\bigg(\frac{a}{l}\bigg)^4-\frac{1}{16}\bigg(\frac{a}{l}\bigg)^5\Bigg]^{-1} -1
\label{eq:drag_wb_faxen2}    
\end{align}
This correlation is in good agreement with experimental data up to $Re_\text{slip} = 0.1$ \citep{Ambari,Takemura}.

\citet{VasseurCox2} analysed the effects of walls located in the outer-region of the flow disturbance produced by the particle on the drag force. The study considered a quiescent flow and used a method of matched asymptotic expansions together with the Oseen approximation. The integral expression obtained for the drag force by solving the outer-region velocity field was numerically evaluated and plotted as a function of $l/L_\text{S}$. Two analytical models valid in the limits of $l \ll L_\text{S}$ and $l\gg L_\text{S}$ were also suggested in the same study. Later, \citet{Takemura} suggested an empirical fit for \citet{VasseurCox2}'s outer-region-based wall-bounded drag coefficient. The model was presented as a function of $l/L_\text{S}$ and considered the numerical values up to $l/L_\text{S} \ll 10$.
\begin{align}
    C_\text{D,2}^\text{wb,out} = 6\pi\bigg(\frac{a}{l}\bigg)\Bigg[\frac{9}{16+11.13(l/L_\text{S})+0.584(l/L_\text{S})^2+0.371(l/L_\text{S})^3}\Bigg]
    \label{eq:drag_wb_takemura_outer}
\end{align}
The net drag coefficient under \citeauthor{Takemura}'s outer-region models is obtained by replacing the $C_\text{D,2}^\text{wb,in}$ by $C_\text{D,2}^\text{wb,out}$ in Eq. (\ref{eq:drag_wb_faxen_cof}) \citep{Takemura}. The theoretical predictions of $C_\text{D,2}^\text{wb,in}$ and $C_\text{D,2}^\text{wb,out}$, given via Eq. (\ref{eq:drag_wb_faxen2}) and Eq. (\ref{eq:drag_wb_takemura_outer}), reduce to zero with increasing separation distance, while the net drag coefficient, $C_\text{D,2}$, reduces to the unbounded drag coefficient value $C_\text{D,2}^\text{ub}$. Note however that as the slip Reynolds number increases, the net drag coefficient predicted using $C_\text{D,2}^\text{wb,out}$ tends to reach the unbounded Stokes limit much faster than predicted via the inner-region $C_\text{D,2}^\text{wb,in}$.

Eqs. (\ref{eq:drag_wb_faxen2}) and (\ref{eq:drag_wb_takemura_outer}) are specific to the inner and outer-regions, respectively, and hence cannot represent the drag across all separation distances. Based on experimental results at $Re_\text{slip} \sim 0.09 - 0.5$, \citet{Takemura} suggested a modification to $C_\text{D,2}^\text{wb}$ as follows:
\begin{align}
     \frac{C_\text{D,2}^\text{wb}}{6 \pi} = \Bigg[1-\bigg(\frac{ C_\text{D,2}^\text{wb,out}}{6 \pi}\frac{l}{a}\bigg)\bigg(\frac{a}{l}\bigg)+\frac{1}{8}\bigg(\frac{a}{l}\bigg)^3-\frac{45}{256}\bigg(\frac{a}{l}\bigg)^4-\frac{1}{16}\bigg(\frac{a}{l}\bigg)^5\Bigg]^{-1} -1
\label{eq:drag_wb_takemura_innerouter}    
\end{align}
In this modification, the $9/16$ coefficient of the inner-region model represented by Eq. (\ref{eq:drag_wb_faxen2}) was replaced by $C_\text{D,2}^\text{wb,out}$ to capture the transition behaviour of the $C_\text{D,2}^\text{wb}$ when a particle shifts from the inner to the outer-region.

For wall-bounded linear shear flows, \citet{Magnaudet1} presented an additional contribution to the drag force due to wall-shear applicable for the inner-region of the disturbed flow at $Re_\gamma \ll 1$. The force was given by:
\begin{align}
    {{F}_\text{D}^\ast} = -\text{sgn} (\gamma)Re_\gamma{}C_\text{D,1}^\text{wb,in}(Re_\gamma,a/l)-\text{sgn} (u_\text{slip})Re_\text{slip}C_\text{D,2}(Re_\text{slip},a/l)
    \label{eq:drag5}    
\end{align}
with the wall-shear based drag coefficient,  $C_\text{D,1}^\text{wb,in}$ given as a function of $(a/l)$ as,
\begin{align}
    C_\text{D,1}^\text{wb,in} = \dfrac{15\pi}{8}\bigg(\dfrac{1}{l^{\ast}}\bigg)^2\Bigg[1+\dfrac{9}{16}\bigg(\dfrac{a}{l}\bigg)\Bigg] 
    \label{eq:drag_cd1_magneduet}    
\end{align}
This function reduces rapidly to zero moving away from the wall \citep{Magnaudet1}. In our previous numerical study, \citet{EkanayakeJFM1} modified this coefficient by including higher order terms of separation distance and introduced an inertial correction for shear, resulting in an expression valid for inner, outer and unbounded regions as:
\begin{align}
    C_\text{D,1} = \dfrac{15\pi}{8}\bigg(\dfrac{a}{l}\bigg)^2\Bigg[1+\dfrac{9}{16}\bigg(\dfrac{a}{l}\bigg)+0.5801\bigg(\dfrac{a}{l}\bigg)^2-3.34\bigg(\dfrac{a}{l}\bigg)^3+4.15\bigg(\dfrac{a}{l}\bigg)^4\Bigg] \notag\\
    +(3.001Re_\gamma^2 -1.025Re_\gamma)
    \label{eq:drag_cd1_nilanka}    
\end{align}
Note that this shear based wall drag creates a negative slip velocity near a wall for force-free particles.

Despite the considerable past research in this area, existing drag models require further work to cover practically relevant moderate inertial ranges. For particles moving in quiescent flows, the influence of finite slip inertial effects on the $C_\text{D,2}$ drag coefficient requires further validation, particularly for $Re_{\text{slip}}<10^{-1}$.  For particles moving in linear shear flows, the influence of both finite slip and shear inertial effects on the overall $C_\text{D}$ drag coefficient also requires further validation, again for the relevant ranges of $Re_{\text{slip}},Re_\gamma<10^{-1}$.

\section{Numerical Simulations}\label{sec:jfm2:NumericalSimulations}
\subsection{Problem Specification}\label{sec:jfm2:Problem Specification}

\begin{figure}
    \begin{subfigure}{0.495\linewidth} 
    \centerline{\includegraphics[width=0.75\linewidth]{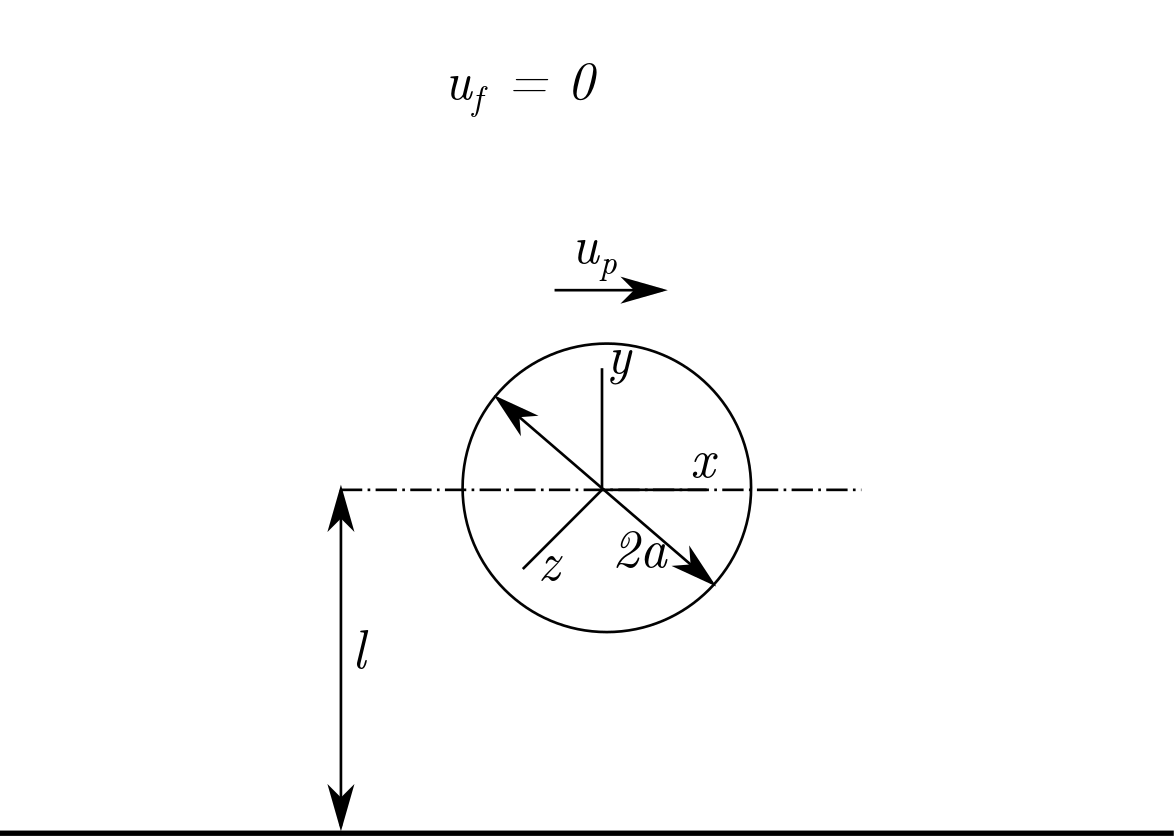}}
    \caption{}
    \label{Diagram1}
    \end{subfigure}
    \begin{subfigure}{0.495\linewidth}
    \centerline{\includegraphics[width=0.75\linewidth]{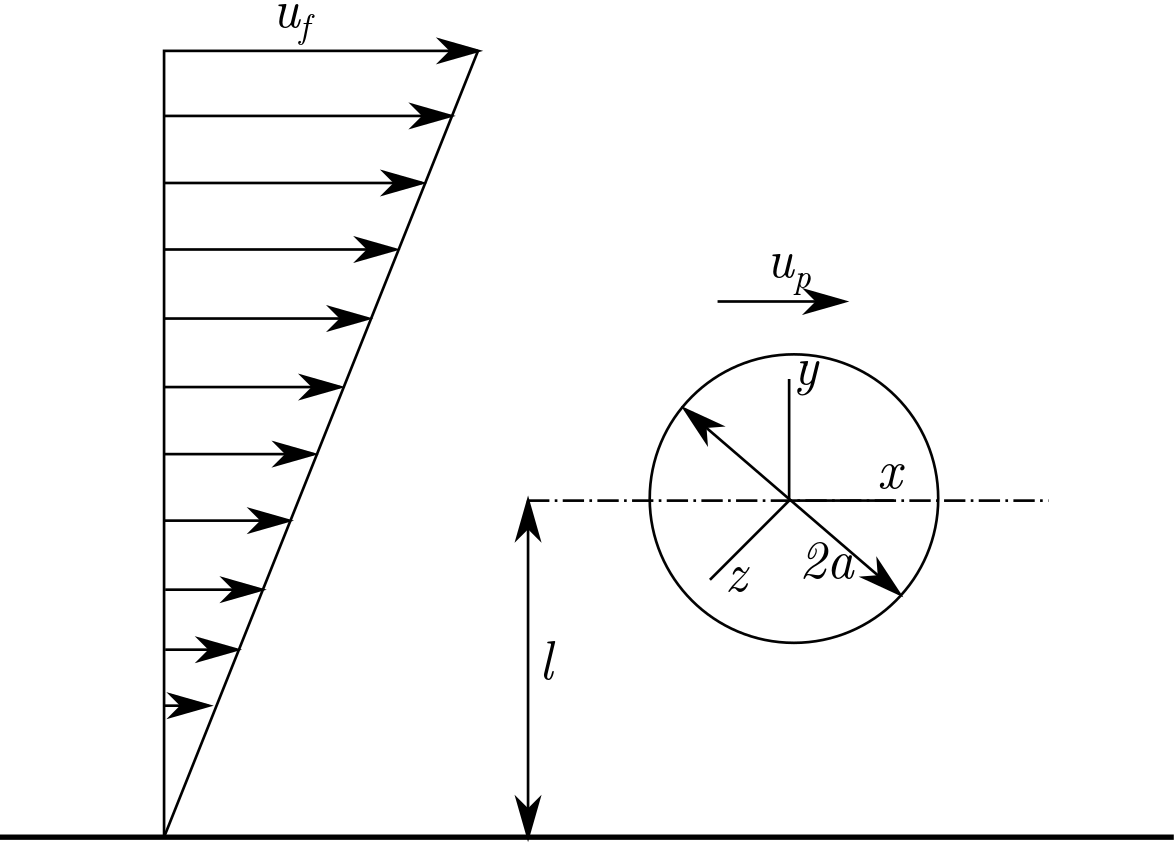}}
    \caption{}
    \label{Diagram2}
    \end{subfigure}
\caption{Schematic of a translating sphere of radius $a$ moving at velocity $u_\text{p}$ in a wall-bounded (a) quiescent flow (b) linear shear flow.}
\label{fig:Schematic}
\end{figure}

The numerical framework of the present investigation is similar to that of \S 2 in \citet{EkanayakeJFM1} except that here the particle moves with a non-zero slip velocity. We consider a rigid sphere of radius $a$ with the origin of the Cartesian coordinate system located at the centre of the sphere. Both quiescent and linear shear fluid flows are considered (figure \ref{fig:Schematic}). For both cases, the particle slip velocity is explicitly set to a known value: 
\begin{align*}
  \boldsymbol{u_{\text{slip}}} =\boldsymbol{u_{\text{p}}}-\boldsymbol{u_{\text{f}}}(y=0) = u_{\text{slip}}\boldsymbol{e_{\text{x}}}
\end{align*}
where ${\boldsymbol{u}_{\text{p}}}=u_\text{p}\boldsymbol{e_{\text{x}}}$ is the particle velocity and ${\boldsymbol{u}_{\text{f}}}$ is the undisturbed fluid. The fluid velocity of the linear shear flow is defined as
\begin{align*}
  \boldsymbol{u_{\text{f}}} &={\gamma}(y+l) \boldsymbol{e_{\text{x}}}
\end{align*}
Here $l$ is the distance of the sphere centre from the wall and  ${\boldsymbol{e}_\text{x}}$ is the coordinate unit vector in $x$ direction. For the quiescent flow cases, $\gamma$ is set to zero. Note that under this formulation the particle is constrained to translate only in the $x$ direction with particle velocity $u_\text{p}$.

A reference frame that moves with the particle \citep{Batchelor67} is employed to solve the steady-state Navier Stokes (N-S) equations:
\begin{subequations}
\begin{gather}
\displaystyle{\boldsymbol{\nabla} \cdot \rho\boldsymbol{u'}} = 0\\
\displaystyle  {\boldsymbol{\nabla} \cdot (\rho\boldsymbol{u'}\boldsymbol{u'}+\boldsymbol{\sigma})} = 0
\end{gather}
\label{NS}%
\end{subequations}
where $\boldsymbol{u'} = \boldsymbol{u} - \boldsymbol{u_{\text{p}}}$ and $\boldsymbol{u}$ is the local fluid velocity. The boundary conditions used in the moving frame of reference are:
\begin{equation}
\boldsymbol{u'}=\left \{ \begin{array}{ll}  
\displaystyle [{\gamma}(y+l)-{u_\text{p}}]\boldsymbol{e_{\text{x}}} \ \ \ \  y=+\infty; y=-l; x,z=\pm\infty \\
\displaystyle \boldsymbol{\omega}\times\boldsymbol{r} \qquad\qquad\ \ \ \ \ {|\boldsymbol{r}|}=a
\end{array}\right.
\label{BC2}
\end{equation}
where $\boldsymbol{r}$ is a radial displacement vector pointing from the sphere centre to the particle surface and $\boldsymbol{\omega}$ is the angular rotation of the particle. The fluid is assumed to be Newtonian with a dynamic viscosity $\mu$ and density $\rho$. The total stress tensor ($\boldsymbol{\sigma}= p\boldsymbol{I}+\boldsymbol{\tau}$) (\citet{BSLBook} sign convention), is computed using the fluid pressure, $p$ and viscous stress tensor, $\boldsymbol{\tau} = -\mu(\boldsymbol{\nabla u'} + \boldsymbol{\nabla u'}^{\text{T}})$.

The forces (${\boldsymbol{F}_{\text{p}}}$) and the torque (${\boldsymbol{T}_{\text{p}}} $) acting on the particle are calculated using the same method provided in \citet{EkanayakeJFM1};
\begin{subequations}
\begin{gather}
    {\boldsymbol{F}_{\text{p}}} = -\int_S \boldsymbol{n}\cdot{\boldsymbol{\sigma}} dS\\
    {\boldsymbol{T}_{\text{p}}} = -\int_S \boldsymbol{r} \times {\boldsymbol{\sigma}}\cdot\boldsymbol{n} dS
\end{gather}
\label{ForceTorque}%
\end{subequations}
where $S$ and $\boldsymbol{n} (=\boldsymbol{\hat{r}})$ are the particle surface area and outward unit normal vector of particle respectively. The drag (${F_\text{D}} = {\boldsymbol{F}_{\text{p}}}\cdot\boldsymbol{e}_{\text{x}}$) and lift (${F_\text{L}}={\boldsymbol{F}_{\text{p}}}\cdot\boldsymbol{e}_{\text{y}} $) are defined as the fluid forces acting on the sphere in the $+x$ and $+y$ directions, respectively. For the non-rotating cases all components of the angular velocity $\boldsymbol{\omega}$ are explicitly set to zero, whereas for the freely-rotating cases the $z$ component of the net torque $\boldsymbol{T}_{\text{p}}$ is explicitly set to zero and the $z$ component of $\boldsymbol{\omega}$ ($\boldsymbol{\omega} \cdot \boldsymbol{e_{\text{z}}} = \omega_\text{p}$) is solved for as an unknown (with other components of $\boldsymbol{\omega}$ set to zero).

The results in the remainder of this study are presented in non-dimensional form (indicated by an asterisk) using length scale $a$, time scale $a/u_\text{slip}$, velocity scale $\gamma a$ and force scale $\mu^2/\rho$.

\subsection{Numerical approach}\label{sec:jfm2:Numerical approach}

The system of equations given in \S\ref{sec:jfm2:Problem Specification} is solved using the finite volume package \textit{arb} \citep{Harvie} over a non-uniform body-fitted structured mesh \citep{EkanayakeJFM1}, generated with \textit{gmsh} \citep{gmsh}.

\subsubsection{Domain Size Dependency}
\begin{table}
\centering
    \begin{tabular}{ccccllllllll}
    \multirow[t]{3}{*}{${l^{\ast}}$} & \multirow[t]{3}{*}{$Re_{\text{slip}}$} & \multicolumn{2}{c}{Domain} & \multicolumn{4}{c}{Non-rotating} & \multicolumn{4}{c}{Freely-rotating} \\\\
    & & & & 
      \multicolumn{2}{c}{Lift} &
      \multicolumn{2}{c}{Drag} &
      \multicolumn{2}{c}{Lift} &
      \multicolumn{2}{c}{Drag}\\\\
    & & ${L^{\ast}}$  & $N_\text{t}$ & $C_{\text{L,3}}$ & 
    $\lvert\delta\%\rvert$ & $-C_{\text{D,2}}$ & $\lvert\delta\%\rvert$ & $C_{\text{L,3}}$ & 
    $\lvert\delta\%\rvert$ & $-C_{\text{D,2}}$ & $\lvert\delta\%\rvert$ \\
    \hline\\
    \multirow{6}{*}{$1.2$}&\multirow{3}{*}{0.001} & $50$ & 158976 & 1.7828 & 1.6284 & -36.831 & 0.1792 & 1.7121 & 1.8196 & -36.786 & 0.1858 \\
    & & $100$ & 189702 & 1.7545 & 0.0138 & -36.766 & 0.0026 & 1.6818 & 0.0165 & -36.719 & 0.0027 \\
    & & $120$ & 200960 & 1.7542 & - & -36.765 & - & 1.6815 & - & -36.718 & -\\\\
    &\multirow{3}{*}{0.1} & $50$ & 158976 & 1.7653 & 1.2595 & -36.849 & 0.1410 & 1.6947 & 1.4308 & -36.804 & 0.1464 \\
    & & $100$ & 189702 & 1.7434 & 0.0149 & -36.798 & 0.0019 & 1.6711 & 0.0175 & -36.751 & 0.0020\\
    & & $120$ & 200960 & 1.7431 & - & -36.797 & - & 1.6708 & - & -36.751 & -\\
    \hline\\
    \multirow{6}{*}{$9.5$}&\multirow{3}{*}{0.001} & $50$ & 236736 & 1.7086 & 5.6979 & -20.222 & 0.7455 & 1.7096 & 5.6488 & -20.222 & 0.7454 \\
    & & $100$ & 280962 & 1.8037 & 0.4493 & -20.084 & 0.0563 & 1.8039 & 0.4444 & -20.084 & 0.0562 \\
    & & $120$ & 296960 & 1.8118 & - & -20.072 & - & 1.8120 & - & -20.072 & - \\\\
    &\multirow{3}{*}{0.1} & $50$ & 236736 & 1.5206 & 2.5743 & -20.362 & 0.5574 & 1.5215 & 2.5324 & -20.362 & 0.5574\\
    & & $100$ & 280962 & 1.5597 & 0.0723 & -20.256 & 0.0348 & 1.5600 & 0.0685 & -20.256 & 0.0348\\
    & & $120$ & 296960 & 1.5608 & - & -20.249 & - & 1.5611 & - & -20.249 & -\\
  \end{tabular}
  \caption{Effect of domain size on drag and lift coefficients for maximum and minimum separation distances (${l^{\ast}} = 1.2$ and $9.5$) and slip Reynolds number ($Re_\text{slip} = 10^{-3}$ and $10^{-1}$) at $Re_{\gamma}=0$. $\delta$ is the percentage error in coefficient, relative to results calculated using the largest domain size (${L^{\ast}}=120$).}
  \label{Domain}
\end{table}
Since the domain and mesh dependency were tested for linear shear flows in our previous study \citep{EkanayakeJFM1}, here we select a quiescent flow. The location of the outer boundaries of the mesh, ${L^{\ast}}$, is first varied to select a suitable domain size such that the lift and drag forces are negligibly affected by this parameter. ${L^{\ast}}$ is increased from $50$ to $120$ and simulations are performed for three selected slip Reynolds numbers; $Re_{\text{slip}} = 10^{-3}, 10^{-2}$ and $10^{-1}$ and for seven wall distances; ${l^{\ast}} = 1.2, 2, 3, 4, 6, 8$ and $9.5$; at $Re_{\gamma}=0$. With increasing ${L^{\ast}}$, the number of mesh points in the domain edge is systematically increased, resulting in $N_{\text{t}}$ total number of cells.

The lift and drag coefficients, $C_\text{L,3}$ and $C_\text{D,2}$ respectively (defined in \S \ref{sec:jfm2:Generalising net lift and drag formula}) are shown in table \ref{Domain} for the minimum and maximum separation distances (${l^{\ast}} = 1.2$ and $9.5$) and minimum and maximum slip Reynolds numbers ($Re_{\text{slip}} = 10^{-3}$ and $10^{-1}$). $\delta$ is the percentage difference of each force coefficient relative to the values obtained using the maximum domain size (${L^{\ast}} = 120$) and is used an indicator of the coefficient accuracy.

For all non-rotating and freely-rotating cases, a domain size of ${L^{\ast}} = 100$ is sufficient to capture the inertial effects responsible for lift results, since $\lvert\delta\rvert \ll 1\%$, with an exception of lift results at ${l^{\ast}} = 9.5$ and $Re_{\text{slip}} = 10^{-3}$. However, even for these conditions, increasing the domain size by $20\%$ (from  ${L^{\ast}} = 100$ to  ${L^{\ast}} = 120$) only results in a change in $C_\text{L,3}$ of less than $1\%$. For smaller separation distances (i.e., ${l^{\ast}} = 1.2$), the reported $\delta$ values are significantly small as the near-wall effects dominate outer boundary effects \citep{Ekanayake}. The $\delta$ values calculated for drag force coefficients are again much less than $1\%$ for the selected domain size of ${L^{\ast}} = 100$ for all separation distances. Hence, a domain size of ${L^{\ast}} = 100$ is used for all simulations in this study.

\subsubsection{Mesh Dependency}
\begin{figure}
\centering
\includegraphics[width=0.8\columnwidth]{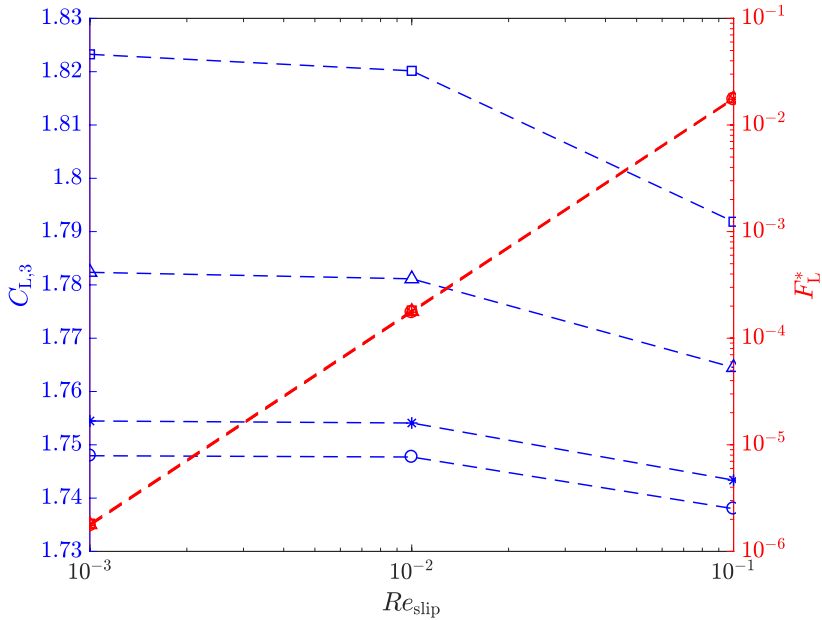}
\caption{Effect of mesh resolution around the sphere on $C_\text{L,3}$ for a non-rotating particle at ${l^{\ast}}=1.2$. $N_\text{p}$ = 15 (\protect\hollowsquaresolidline); 20 (\protect\hollowtrainglesolidline); 25 (\protect\hollowstarsolidline); 30 (\protect\hollowcirclesolidline).} 
\label{Mesh1}
\end{figure}
In this section the effect of mesh resolution within the boundary layers surrounding the sphere is examined. The number of cells on the sphere surface and the number of inflation layers around the sphere are adjusted systematically by varying the number of mesh points, $N_\text{p}$, on each curved side length of a cubed-sphere. 

Figure \ref{Mesh1} shows the variation of lift coefficients and non-dimensionalised lift forces as a function of $Re_\text{slip}$ for four mesh refinement levels. The simulations are performed for the smallest separation distance ${l^{\ast}}=1.2$ with a domain size of ${L^{\ast}}=100$. As $Re_{\text{slip}}$ approaches zero, $F_\text{L}^{\ast}$ reduces to zero while $C_\text{L,3}$ asymptotes to different finite values. While results for $F_\text{L}^{\ast}$ appear to be independent of mesh refinement, $C_\text{L,3}$ exhibits considerable variation with mesh refinement, particularly for low $Re_{\text{slip}}$ values. This relative difference in $C_\text{L,3}$ decreases as $N_\text{p}$ increases. For example, at the lowest $Re_{\text{slip}}$ value, $C_\text{L,3}$ changed by $3.77\%$ as $N_\text{p}$ is increased from $15$ to $25$, but changes by only $0.37\%$ as $N_\text{p}$ is increased from $25$ to $30$. Noting the significant increase of the total cell count $N_\text{t}$ from 158,976 to 310,500 with increasing $N_\text{p}$ from $25$ to $30$ and by considering the computational memory requirements, we employed the mesh with $N_\text{p} = 25$ for the remainder of the study.

\section{Numerical Results and Force Correlations}\label{sec:jfm2:Results and Discussion}
In this section, we first provide our generalised lift and drag force definitions (\S\ref{sec:jfm2:Generalising net lift and drag formula}), and then develop new force correlations based on the numerical results for quiescent flows (\S\ref{sec:jfm2:uniformflow}) and linear shear flows (\S \ref{sec:jfm2:constantepsi}).


\subsection{Lift and drag model definitions}\label{sec:jfm2:Generalising net lift and drag formula}

Here we use the definition of \citet{EkanayakeJFM1}, applicable for inner, outer and unbounded regions, to present the lift force in a linear shear flow for finite slip and shear conditions,
\begin{gather}
     {F_\text{L}^{\ast}} = C_\text{L,1}Re_{\gamma}^2+\text{sgn}({\gamma}^{\ast})C_\text{L,2}Re_{\gamma}Re_{\text{slip}}+C_\text{L,3}Re_{\text{slip}}^2
     \label{eq:jfm2:netlift_allregion_0}
\end{gather}
The first and last terms of Eq. (\ref{eq:jfm2:netlift_allregion_0}) are defined by the lift forces in a linear shear flow in the absence of slip ($Re_\text{slip} = 0$) and in a quiescent flow in the absence of shear ($Re_\gamma = 0$), respectively. The remaining term captures the remaining lift contributions in the presence of both slip and shear. The lift coefficients $C_\text{L,1}$, $C_\text{L,2}$ and $C_\text{L,3}$ in Eq. (\ref{eq:jfm2:netlift_allregion_0}) are defined to be functions of shear, shear and slip, and slip, respectively, as well as the wall distance. The unambiguous definitions for the three coefficients allows Eq. (\ref{eq:jfm2:netlift_allregion_0}) to be a valid representation of the lift force at any separation distance.

Similarly, the drag force for finite slip in a linear shear flow is defined as \citep{EkanayakeJFM1},
\begin{gather}
    {F_\text{D}^{\ast}} = -\text{sgn}(\gamma)C_\text{D,1}Re_{\gamma} -\text{sgn}(u_\text{slip})C_\text{D,2}Re_{\text{slip}}
    \label{eq:jfm2:netdrag_allregion_0}
\end{gather}
The first term in Eq. (\ref{eq:jfm2:netdrag_allregion_0}) is associated with the drag force in a linear shear flow in the absence of slip ($Re_\text{slip} = 0$) while the second term captures the remaining drag contributions in the presence of both slip and shear. Both coefficients are functions of wall distance, with $C_\text{D,1}$ a function of shear, and $C_\text{D,2}$ a function of both slip and shear. Both force coefficients, $C_\text{D,1}$ and $C_\text{D,2}$ defined in this equation are valid for arbitrary separation distances (inner, outer, unbounded regions).

In the present study, we investigate forces on a particle under finite slip conditions, and hence, provide new correlations for lift coefficients $C_\text{L,2}$ and $C_\text{L,3}$. In addition, the most suitable correlation for $C_\text{D,2}$ under linear shear flow conditions is determined for finite slip and shear Reynolds numbers. In the remainder of this study, unless stated otherwise, the zero-slip lift and drag force coefficients ($C_\text{L,1}$ and $C_\text{D,1}$, respectively) are evaluated using the correlations proposed by \citet{EkanayakeJFM1} that are valid for all separation distances $l^{\ast} \geq 1.2$.

\subsection{Particle translating in a quiescent flow}\label{sec:jfm2:uniformflow}

\subsubsection{Lift force}
\begin{figure}
\begin{subfigure}{\linewidth}
\centerline{\includegraphics[width=0.80\linewidth]{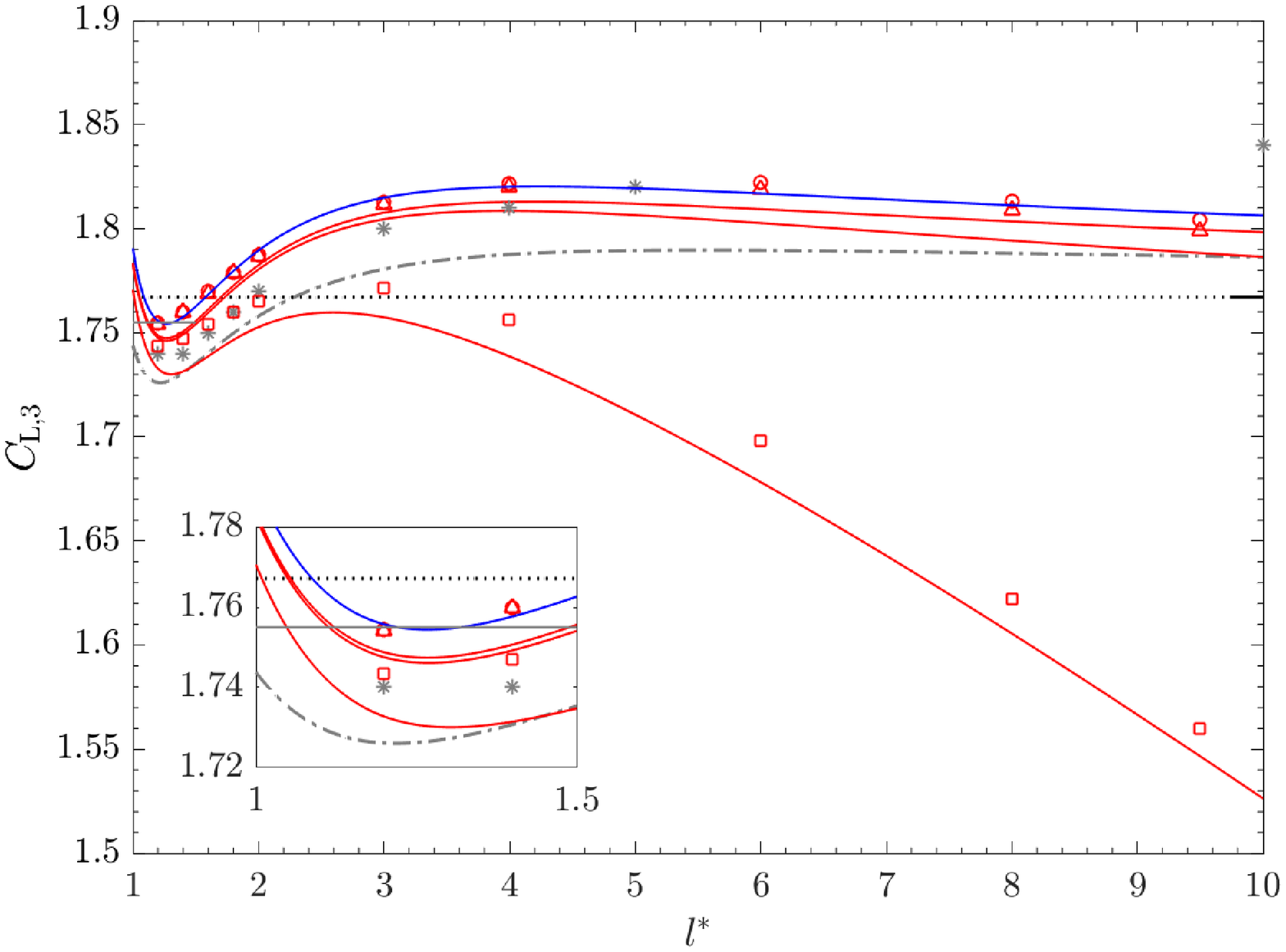}}
\caption{non-rotating}
\label{fig:Lift_shear0_omega0_fit}
\end{subfigure}
\begin{subfigure}{\linewidth}
\centerline{\includegraphics[width=0.80\linewidth]{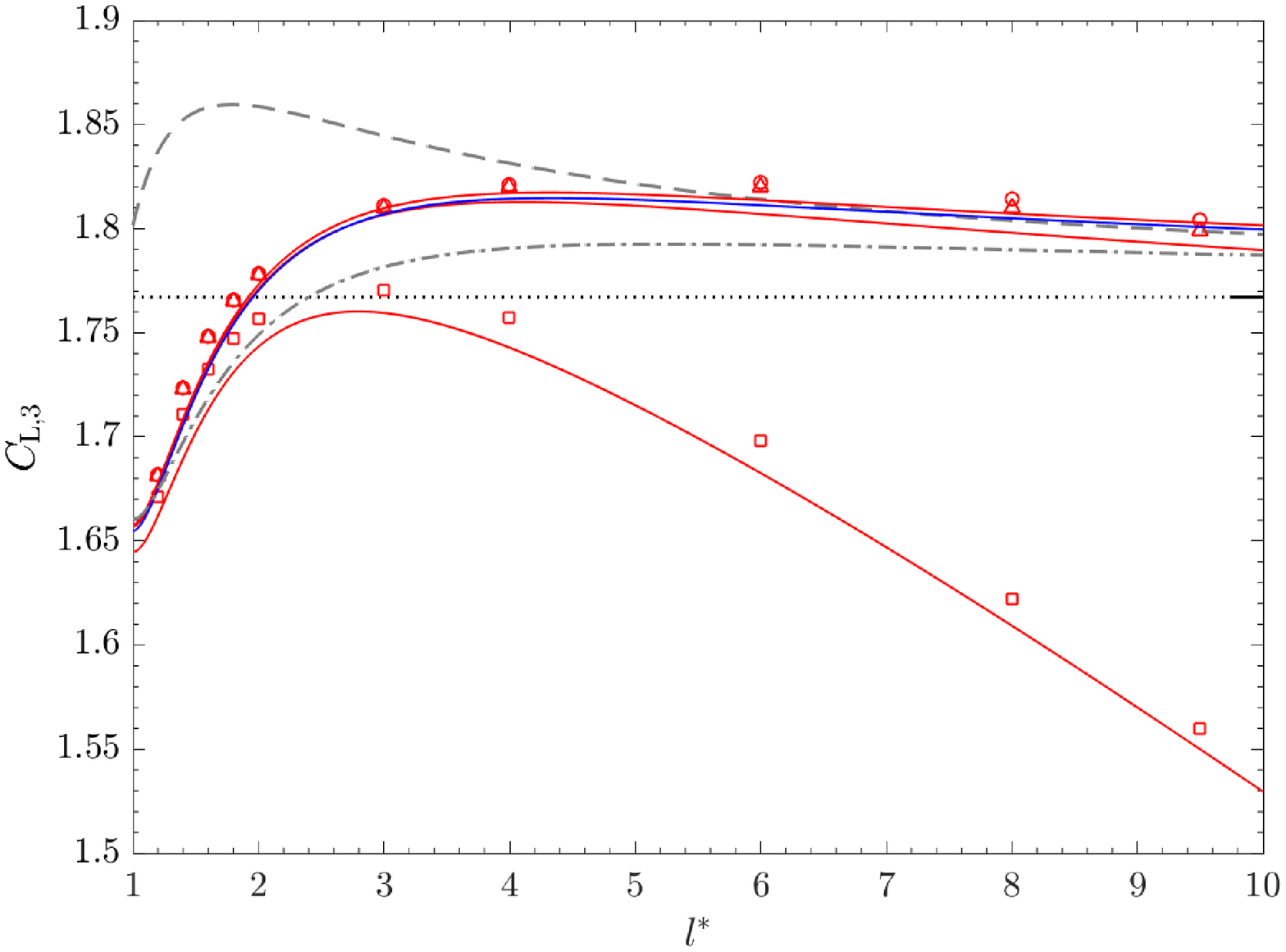}}
\caption{freely-rotating}
\label{fig:Lift_shear0_torque0_fit}
\end{subfigure}
\caption{Lift coefficient ($C_{\text{L,3}}$) for different shear Reynolds number as a function of non-dimensional separation distance (${l^{\ast}}$) for (a) non-rotating and (b) freely-rotating spheres. Simulations: $Re_\text{slip} = 10^{-3}$ (\protect\hollowredcircle), $10^{-2}$ (\protect\hollowtrainglered) and $10^{-1}$ (\protect\hollowsquarered). Numerical predictions by \citet{Fischer87} that included small inertial effects at $Re_\gamma \ll 1$ (\protect\blackstar). Analytical predictions of \citet{CoxHsu} (\protect\blackdottedline, Eqs. \ref{Cox1},\ref{Cox2}), \citet{CherukatMcLaughlin} (\protect\Graydashdottedline, Eqs. \ref{CM1}, \ref{CM2}), \citet{Krishnan} (\protect\Grayline, Eqs. \ref{KL1}) and \citet{Magnaudet1} (\protect\Graydashedline, Eq. \ref{MD1}) . Present numerical fit for inner-region (\protect\blueline, Eqs. \ref{eq:lift0shear_0omega}, \ref{eq:lift0shear_0toque}). Present numerical fit for all regions (\protect\redline, Eqs. \ref{eq:lift0shear_0omega_inout},\ref{eq:lift0shear_0toque_inout})}
\label{fig:Lift_shear0_fit}
\end{figure}

In figure \ref{fig:Lift_shear0_fit}, the lift coefficients $C_\text{L,3}$ computed for both non-rotating and a freely-rotating particles in a quiescent flow are plotted as a function of non-dimensional separation distance (${l^{\ast}}$). The numerical results are compared against the available inner-region correlations listed in table \ref{table:InnerLiftSummary} that are valid for $Re_{\text{slip}} \ll 1$. In general, the lift coefficient values predicted via most of the analytical solutions slightly underestimate the numerically computed lift forces, particularly for $Re_{\text{slip}} < 10^{-2}$ near the wall. For example, the lowest slip Reynolds number ($Re_{\text{slip}} = 10^{-3}$) simulation conducted at the smallest distance to the wall (${l^{\ast}} = 1.2$) gives a $C_{\text{L,3}}$ of $1.755$ ($1.682$) for a non-rotating (freely-rotating) particle, which is $\sim 1.68\%$ ($0.48\%$) higher than the asymptotic value of $1.726$ ($1.674$) predicted for a non-rotating (freely-rotating) particle at ${l^{\ast}} = 1.2$ \citep{CherukatMcLaughlin}. The \citet{CoxHsu} first order lift expression, which does not account for the finite particle size, produces a lift coefficient independent of ${l^{\ast}}$ for both non-rotating and a freely-rotating particles. The value agrees to a certain extent with the present numerical and other theoretical predictions, but is not particularly accurate near the wall. When a particle is almost in contact with the wall ($l^{\ast} = 1$), the \citet{Krishnan} study gives a $C_\text{L,3}$ of $1.755$ for a non-rotating particle, which is reasonably consistent with the present numerical results obtained at $l^{\ast} = 1.2$. For a freely-rotating particle at $l^{\ast} = 1$, \citet{Krishnan} also predict a value of $C_\text{L,3} = 0.236$, which is nearly an order of magnitude less than the non-rotating $C_\text{L,3}$ value, and is far from ours and the other inner-region model predictions in this region. Although the rotation of the sphere acts to decrease this lift for a particle near the wall, the reason for the significant deviation between these two theoretical analyses is not clear. For a freely-rotating particle the \citet{Magnaudet1} lift correlation predicts a lift coefficient which is larger than the available asymptotic inner-region theories for $Re_\text{slip} \ll 1$ (Figure \ref{fig:Lift_shear0_torque0_fit}). A similar over-prediction is observed for $C_\text{L,1}$ near the wall (see \citet{EkanayakeJFM1}) which may be caused by the neglect of higher order separation distance terms ($\mathcal{O}(1/{l^{\ast}}) > 2$) that are significant when in the vicinity of the wall.

For the non-rotating case (figure \ref{fig:Lift_shear0_omega0_fit}), the lift results are also compared with numerical predictions based on a Boundary Element Method (BEM) which included small inertial effects ($Re_\gamma \ll 1$) \citep{Fischer87}. The computed lift coefficient values for the lowest slip Reynolds number ($Re_\text{slip} = 10^{-3}$) are consistent with the BEM results for small separation distances (figure \ref{fig:Lift_shear0_omega0_fit}). However, a considerable difference between the present results and BEM predictions is apparent at larger separation distances (${l^{\ast}} \sim 10$). This could be possibly due to insufficient numerical accuracy of a Gauss-Legendre product formula used by \citet{Fischer87}, as suggested by \citet{Shi2}. However, to our knowledge there is no numerical data available for a freely-rotating particle at these low slip Reynolds numbers for additional verification. We also note from our domain dependence study that the errors in our numerical simulations are also highest at these large $l^{\ast}$ and small $Re_\text{slip}$ values. 

As $Re_{\text{slip}}$ increases, the computed lift coefficients deviate significantly from the asymptotic inner-region correlations as inertial effects become significant, in both the non-rotating and freely-rotating cases. Although most of our numerical data are well within the inner-region, the computed lift force coefficient decreases with $Re_\text{slip}$ number in contrast to the inner-region models which predict lift coefficients that are independent of $Re_\text{slip}$. The discrepancy between simulation and theory arises as the force expansion used in the inner-region models does not satisfy the boundary conditions at large distances from the wall. With increasing slip velocity and separation distance, the walls move to the outer-region, and the inner-region-based theoretical lift models then fail to capture the lift coefficient variation. 

\begin{figure}
\centering
\includegraphics[width=0.8\columnwidth]{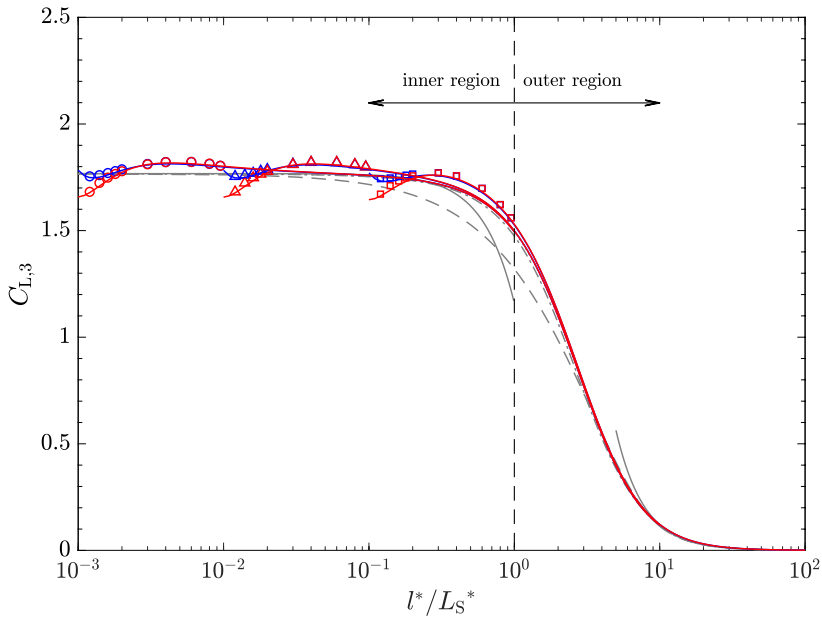}
\caption{Lift coefficient ($C_{\text{L,3}}$) for slip Reynolds number values of $Re_\text{slip} = 10^{-3}$ (\protect\hollowcircle), $10^{-2}$ (\protect\hollowtraingleblack) and $10^{-1}$ (\protect\hollowsquareblack) as a function of separation distance non-dimensionalised by Stokes length scale ($l^{\ast}/{L_{\text{S}}}^{\ast}$). Red and blue symbols are for freely-rotating and non-rotating particles, respectively. Asymptotic models for $Re_\text{slip} \ll 1$ by \citet{VasseurCox2} (\protect\Grayline, Eq. \ref{VC2}), \citet{Takemura2003} (\protect\Graydashedline, Eq. \ref{TM2003}), \citet{Takemura} (\protect\Graydottedline, Eq. \ref{TM2004}) and \citet{Shi2} (\protect\Graydashdottedline, Eq. \ref{ShiCL3}). Present numerical fit for non-rotating particles (\protect\blueline, Eq. \ref{eq:lift0shear_0omega_inout}) and freely rotating particles (\protect\redline, Eq. \ref{eq:lift0shear_0toque_inout})}
\label{fig:Lift_shear0_outer}
\end{figure}

The numerical lift coefficient values and the outer-region asymptotic predictions valid for $Re_\text{slip} \ll 1$ are plotted in figure \ref{fig:Lift_shear0_outer} as a function of $l^{\ast}/{L_\text{S}}^{\ast}$. Although all the computed results are in inner-region, the coefficients closer to the outer boundary ($l^{\ast}/{L_\text{S}}^{\ast} \sim 1$), particularly at $Re_\text{slip} = 10^{-1}$, trend towards the outer-region theoretical predictions. In this boundary limit, the difference between a non-rotating and freely-rotating lift coefficient value is less significant and the lift results are consistent with outer-region theory \citep{VasseurCox2}. Since the outer-region asymptotic models are strictly valid for $l^{\ast}/{L_\text{S}}^{\ast} \gg 1$, variations in the lift correlation observed to occur in the inner-region are thus not captured. This further highlights the need for a model capable of capturing both inner and outer-region slip-lift behaviours simultaneously.

Based on our lift results obtained for the lowest slip Reynolds number ($Re_\text{slip} = 10^{-3}$), we first propose a numerical fit for the wall-slip lift force in the inner-region as a function of separation distance ($l^{\ast}$). The proposed correlations are given by:
\begin{subequations}
\begin{align}
    C_\text{L,3}^\text{wb,in}=1.774 + 0.4353\bigg(\dfrac{1}{l^{\ast}}\bigg) -1.198\bigg(\dfrac{1}{l^{\ast}}\bigg)^2 + 0.7792\bigg(\dfrac{1}{l^{\ast}}\bigg)^3
    \label{eq:lift0shear_0omega}\\
    \intertext{for a non-rotating particle and}
    C_\text{L,3}^\text{wb,in}=1.764 + 0.4757\bigg(\dfrac{1}{l^{\ast}}\bigg) -1.268\bigg(\dfrac{1}{l^{\ast}}\bigg)^2 + 0.683\bigg(\dfrac{1}{l^{\ast}}\bigg)^3
    \label{eq:lift0shear_0toque}
\end{align}
\label{eq:jfm2:lift0shear}%
\end{subequations}
for a freely-rotating particle. Figure \ref{fig:Lift_shear0_fit} indicates that there is no significant variation of the numerical results for $C_\text{L,3}$ when the slip Reynolds number increases from $Re_\text{slip} = 10^{-3}$ to $10^{-2}$. This behaviour suggests that the proposed inner-region-based correlations, shown in figure \ref{fig:Lift_shear0_fit}, can be used for very small slip Reynolds numbers ($Re_\text{slip} \ll 1$) as the numerical lift results are almost independent of slip for $Re_\text{slip}<10^{-2}$. 

Next, by replacing the constant in the proposed inner-region lift model (Eq. \ref{eq:jfm2:lift0shear}) with \citet{Takemura}'s outer-region model (Eq. \ref{TM2004}), the following correlation is proposed to account for the inertial dependency over all wall separation distances as:
\begin{subequations}
\begin{align}
    C_\text{L,3}= C_\text{L,3}^\text{wb,out} + 0.4353\bigg(\dfrac{1}{l^{\ast}}\bigg) -1.198\bigg(\dfrac{1}{l^{\ast}}\bigg)^2 + 0.7792\bigg(\dfrac{1}{l^{\ast}}\bigg)^3
    \label{eq:lift0shear_0omega_inout}\\
    \intertext{for a non-rotating particle and}
    C_\text{L,3}=C_\text{L,3}^\text{wb,out}  + 0.4757\bigg(\dfrac{1}{l^{\ast}}\bigg) -1.268\bigg(\dfrac{1}{l^{\ast}}\bigg)^2 + 0.683\bigg(\dfrac{1}{l^{\ast}}\bigg)^3
    \label{eq:lift0shear_0toque_inout}\\
    \intertext{for a freely-rotating particle. Here, \citet{Takemura}'s outer-region model (Eq. \ref{TM2004}) gives a definition for $C_\text{L,3}^\text{wb,out}$ as}
    C_\text{L,3}^\text{wb,out} = 
    \frac{18\pi}{32+2\bigg(\frac{l^{\ast}}{L_\text{S}^{\ast}}\bigg)+3.8\bigg(\frac{l^{\ast}}{L_\text{S}^{\ast}}\bigg)^2+0.049\bigg(\frac{l\ast}{L_\text{S}\ast}\bigg)^3}\notag
\end{align}
\label{eq:lift0shear_inout}%
\end{subequations}
These correlations are also plotted in figures \ref{fig:Lift_shear0_fit} and \ref{fig:Lift_shear0_outer}, demonstrating that inertial effects are accurately predicted as a particle moves from the inner to outer-region, and the correct limit of $C_\text{L,3} \rightarrow 0$ for $l^{\ast}/L_\text{S}^{\ast} \rightarrow \infty$ is achieved. When a particle is very close to the wall for the smallest Reynolds number, i.e., $l^{\ast}/L_\text{S}^{\ast} \rightarrow 0$ and $l^\ast = 1$, Eq. \ref{eq:lift0shear_0omega_inout} (Eq. \ref{eq:lift0shear_0toque_inout}) calculates the lift coefficient as 1.784 (1.659) for a non-rotating (freely-rotating) particle, a value that is just $2.3\% (0.1\%)$ higher (lower) than the asymptotic inner-region result of \citet{CherukatMcLaughlin} (\citet{CherukatMcLaughlinCorrection}). 

\subsubsection{Drag Force}

\begin{figure}
\centerline{\includegraphics[width=0.85\textwidth]{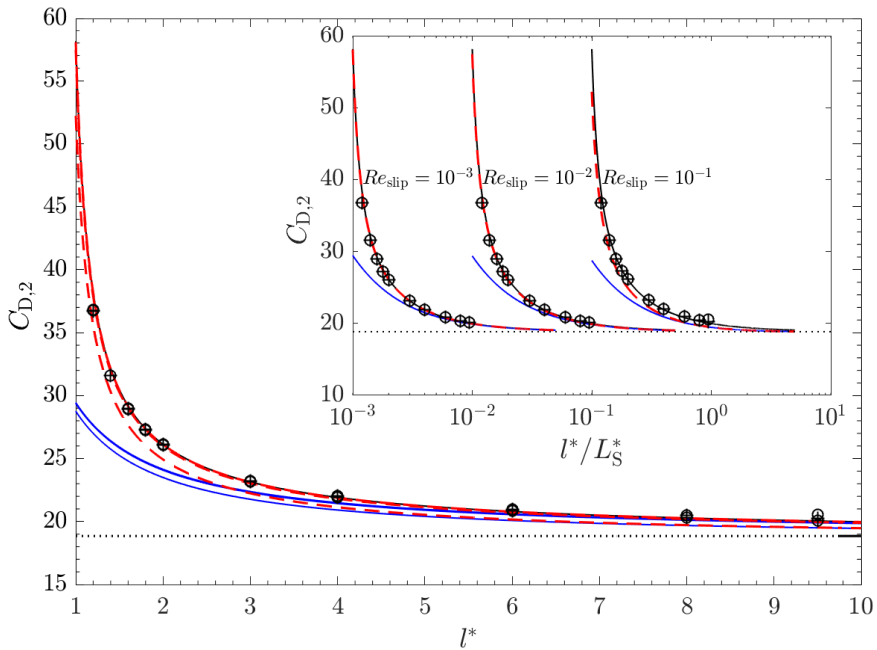}}
\caption{Drag force coefficient of a spherical particle in the absence of shear ($C_{\text{D,2}}$). Simulations: Non-rotating (\protect\hollowcircle), freely-rotating (+). Analytical prediction: Faxen inner-region correlation \citep{Happel} (\protect\blackline, Eq.\ref{eq:drag_wb_faxen2}), \citet{Takemura} outer-region correlation (\protect\blueline, Eq. \ref{eq:drag_wb_takemura_outer}) and \citet{Takemura} inner-outer-region correlation (\protect\reddashline, Eq. \ref{eq:drag_wb_takemura_innerouter}). Unbounded Stokes drag (\protect\blackdottedline, Eq. \ref{IC_Stokes})}
\label{fig:Dragcd_shear0}
\end{figure}

In this section we validate existing drag models using the numerical data obtained for quiescent flows. Under these conditions there is no shear and hence $C_\text{D,1}=0$. Figure \ref{fig:Dragcd_shear0} shows the variation in net drag coefficient, $C_\text{D,2}$, as a function of dimensionless separation distance ($l^{\ast}$) for both non-rotating and freely-rotating particles. The inset in the figure shows the variation as a function of separation distance normalised by Stokes length scale ($l^{\ast}/L_\text{S}^{\ast}$). The results are compared against the wall-bounded inner-region analytical correlation of Faxen \citep{Happel} (Eq. \ref{eq:drag_wb_faxen2}) and the outer-region empirical drag model of \citet{Takemura} (Eq. \ref{eq:drag_wb_takemura_outer}). The predictions of the \citet{Takemura} inner-outer based theoretical model given by Eq. \ref{eq:drag_wb_takemura_innerouter} for low slip Reynolds numbers are also shown.

For each of the slip Reynolds numbers, the highest numerical value for $C_\text{D,2}$ is reported when the particle is close to the wall, and with increasing wall distance, these coefficients asymptote to the unbounded Stokes limit ($C_\text{D,2}=6\pi$) for both non-rotating and freely-rotating particles. The effect of rotation on the drag force is hardly discernible for all separation distances. As indicated in figure \ref{fig:Dragcd_shear0}, no inertial dependency is observed in the computed drag coefficients for all tested $Re_\text{slip}$ values. The numerical drag results are in good agreement with the analytical inner-region correlation, Eq. (\ref{eq:drag_wb_faxen2}), noting that all the numerical results are inside the region $l^{\ast}/L_\text{S}^{\ast} < 1$ (inset of figure \ref{fig:Dragcd_shear0}). However, even in the outer boundary limit (when $l^{\ast}/L_\text{S}^{\ast} \sim 1$), the results given for $Re_{\text{slip}}  = 10^{-1}$ do not significantly deviate from the inner-region predictions (Eq. \ref{eq:drag_wb_faxen2}) or follow the transition behaviour predicted by \citet{Takemura} in Eq. (\ref{eq:drag_wb_takemura_outer}). Note that Eq. (\ref{eq:drag_wb_takemura_outer}) was originally validated for relatively large slip Reynolds numbers ($0.09 \leq Re_{\text{slip}} \leq 0.5$), compared to our simulated slip range. Hence, we conclude that the correlation given by Eq. (\ref{eq:drag_wb_faxen2}) is valid up to $Re_{\text{slip}} = 10^{-1}$ for both rotating and freely-rotating particles without requiring correction any further for inertial effects.

\subsection{Particle translating with a finite slip in a shear flow}\label{sec:jfm2:constantepsi}

In this section, we build on the previous results by analysing the lift and drag force acting on a spherical particle moving parallel to a wall with a finite slip in a linear shear field. We again use the force correlations given in \S\ref{sec:jfm2:Generalising net lift and drag formula} to express the results in terms of lift and drag coefficients. Both $Re_\text{slip}$ and $Re_\gamma$ are varied systematically covering a range from $10^{-3}$ to $10^{3}$ (corresponding to $0.32 < \epsilon < 326$ and $<10^{-2} < |\gamma^{\ast}| <10^{2}$). Both positive and negative slip velocities are considered.

\subsubsection{Lift force}
\begin{figure}
    \begin{subfigure}{\linewidth}
    \centerline{\includegraphics[width=0.85\textwidth]{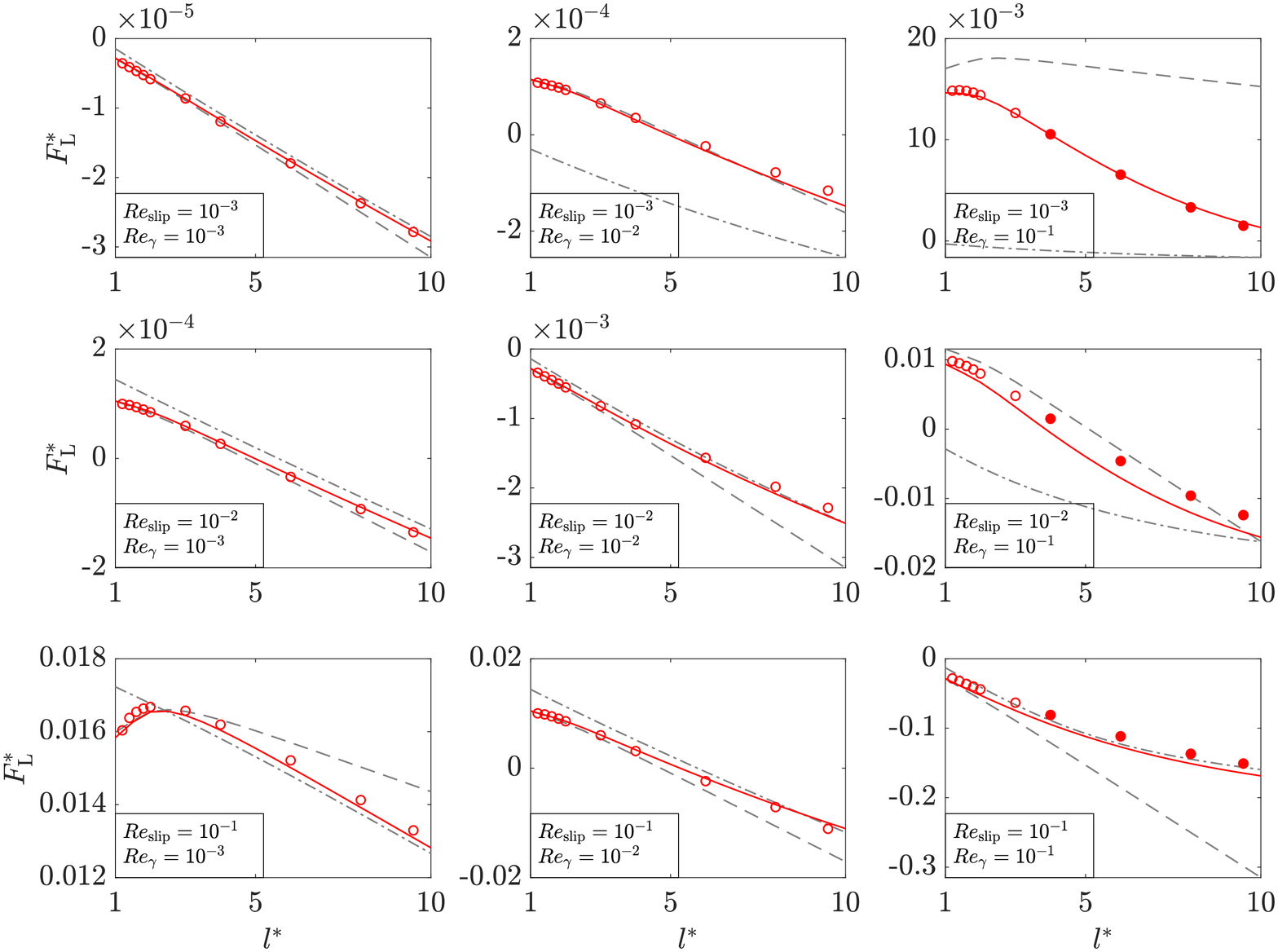}}
    \caption{Leading particle in a positive shear field (${\gamma}^{\ast} > 0$)}
    \label{fig:lift_constantepsi_positive}
    \end{subfigure}
    \begin{subfigure}{\linewidth}
    \medskip
    \centerline{\includegraphics[width=0.85\textwidth]{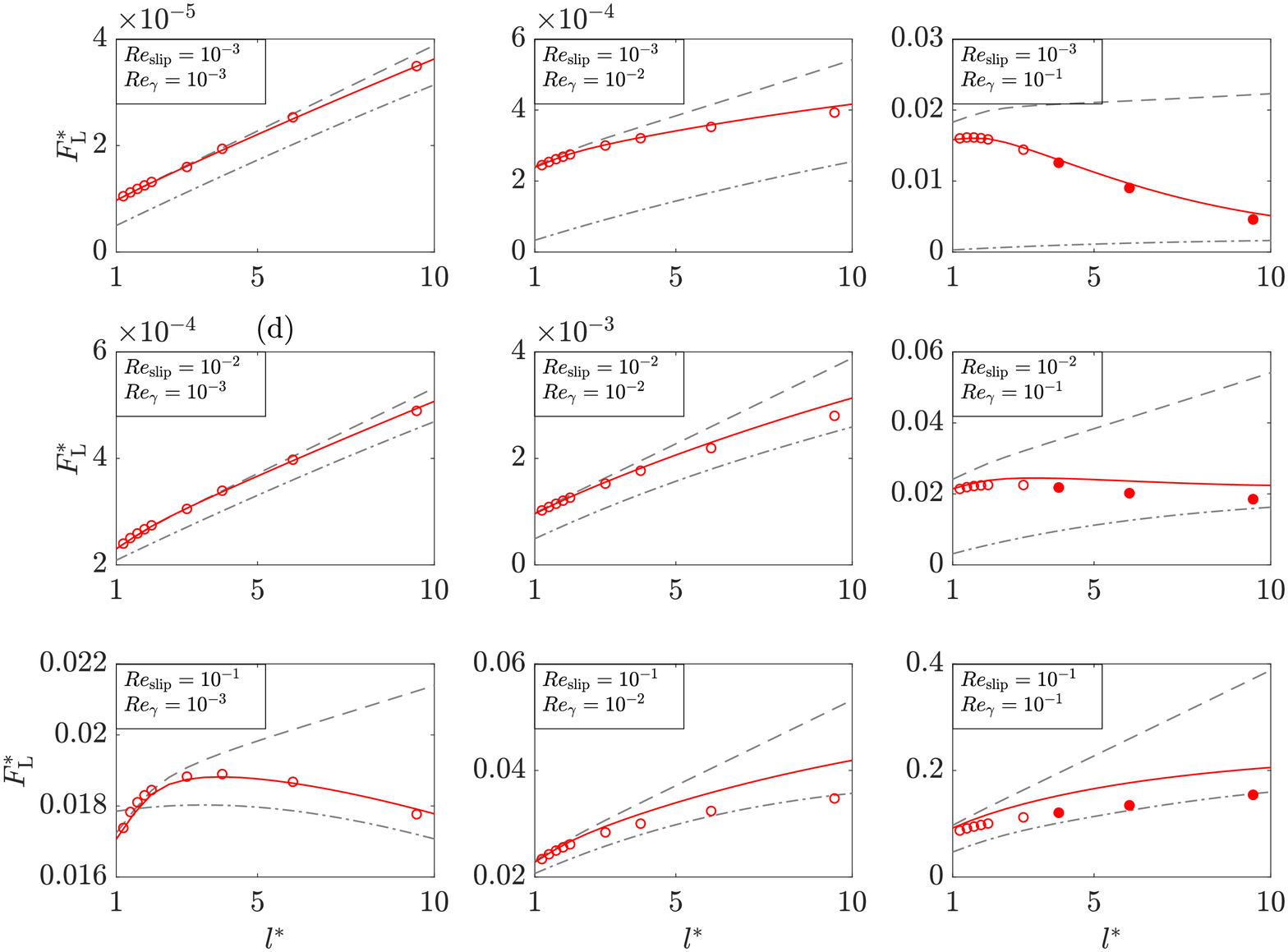}}
    \caption{Lagging particle in a positive shear field (${\gamma}^{\ast} < 0$)}
    \label{fig:lift_constantepsi_negative}
    \end{subfigure}
\caption{Non-dimensional lift force ($F_{\text{L}}^{\ast}$) of a freely-rotating particle. $Re_\gamma$ and $Re_\text{slip}$ increase in the order of $10^{-3}, 10^{-2}, 10^{-1}$ from left to right and top to bottom respectively. Simulations: inner-region (\protect\hollowredcircle) and outer-region ($\textcolor{red}{\bullet}$) for each subplot. Analytical predictions: inner-region model of \citet{CherukatMcLaughlinCorrection} (\protect\Graydashedline, Eq. \ref{eq:liftinner}), outer-region model of  \citet{TakemuraMagnaudet1} (\protect\Graydashdottedline, Eq. \ref{eq:slipshearlift-outercl2cl3}). Present model prediction (\protect\redline, Eq. \ref{eq:newmodel_test})}
\label{fig:lift_constantepsi}
\end{figure}

Figure \ref{fig:lift_constantepsi} shows the variation in dimensionless lift force ($F_\text{L}^{\ast}$) for a freely-rotating particle as a function of $l^{\ast}$ for different combinations of slip and shear Reynolds numbers. Figure \ref{fig:lift_constantepsi_positive} and figure \ref{fig:lift_constantepsi_negative} present results for a leading ($u_\text{slip} > 0$) and a lagging particle ($u_\text{slip}<0$) respectively, in a positive shear field ($\gamma > 0$). The numerical results are compared against the available inner and outer-region lift models that are valid for $Re_\gamma, Re_\text{slip} \ll 1$. Here the inner-region model of \citet{CherukatMcLaughlinCorrection} given by Eq. (\ref{eq:liftinner}), and the outer-region model of \citet{TakemuraMagnaudet1} given by Eq. (\ref{eq:slipshearlift-outercl2cl3}) are included for comparison.

For $Re_\text{slip}, Re_{\gamma} < 10^{-2}$, the numerically computed lift forces in the region close to the wall (${l^{\ast}} < 5$) agree reasonably well with the asymptotic values predicted by the inner-region model \citep{CherukatMcLaughlinCorrection} for both positive and negative slip velocities. As $Re_\text{slip}$ and $Re_{\gamma}$ increase and inertial effects become more significant, the computed lift coefficients deviate significantly from the inner-region theoretical values. With increasing slip, shear and separation distance, the walls move to the outer-region ($l^{\ast}>\text{min}({L_\text{G}}^{\ast},{L_\text{S}}^{\ast})$) and unsurprisingly, the inner-region-based models fail to capture the lift coefficient variations accurately. At $Re_\text{slip}$ and $Re_\gamma>10^{-2}$ and $l^{\ast} \gtrsim 5$, the computed results coincide reasonably well with values predicted by the \citet{TakemuraMagnaudet1} outer-region correlation. However, for the largest shear Reynolds numbers and smallest slip Reynolds numbers (i.e., results for $Re_\gamma > 10^{-2}, Re_\text{slip} < 10^{-1}$), the existing outer-region model significantly underestimates the simulated lift results. 

\begin{figure}
    \begin{subfigure}{\linewidth}
    \centerline{\includegraphics[width=0.8\textwidth]{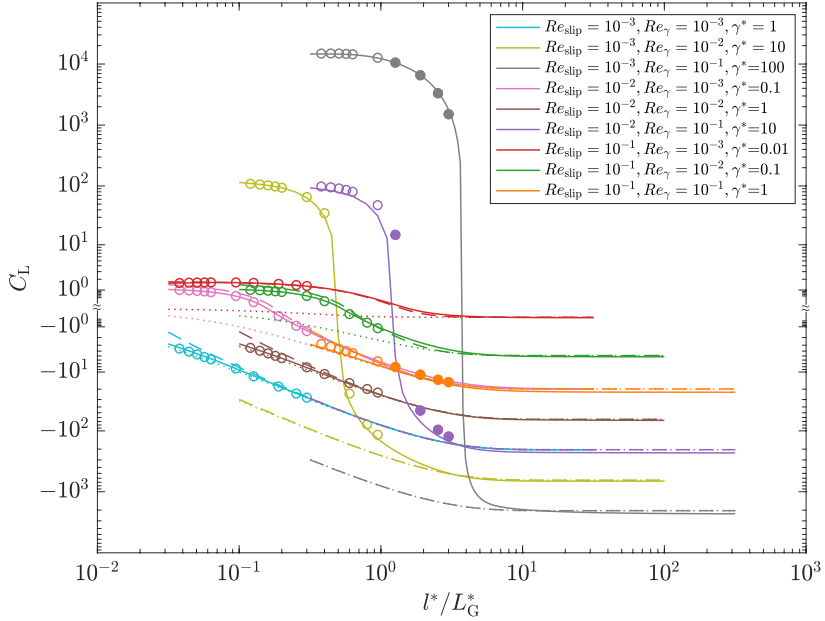}}
    \caption{Leading particle in a positive shear field (${\gamma}^{\ast} > 0$)}
    \label{fig:lift_constantepsi_positive_out}
    \end{subfigure}
    \begin{subfigure}{\linewidth}
    \medskip
    \centerline{\includegraphics[width=0.8\textwidth]{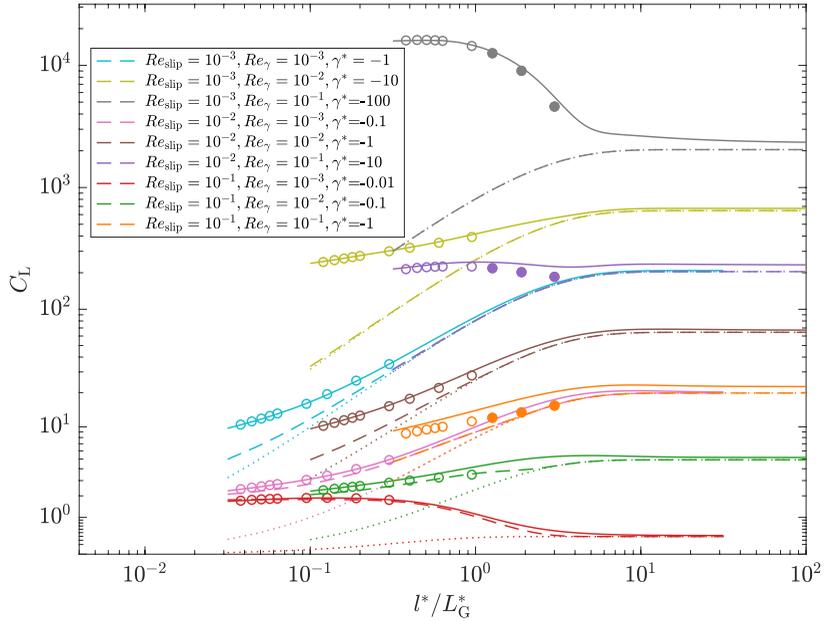}}
    \caption{Lagging particle in a positive shear field (${\gamma}^{\ast} < 0$)}
    \label{fig:lift_constantepsi_negative_out}
    \end{subfigure}
\caption{Lift force coefficient ($C_{\text{L}}=F_{\text{L}}^{\ast}/Re_\text{slip}^2$) of a freely-rotating particle as a function of separation distance non-dimensionalised by Saffman length scale ($l^{\ast}/{L_{\text{G}}}^{\ast}$). Simulations: inner-region (\protect\hollowcircle) and outer-region ($\bullet$). Analytical outer-region correlation of \citet{TakemuraMagnaudet1} (\protect\blackdashedline, Eq. \ref{eq:slipshearlift-outercl2cl3}) and (\protect\blackdottedline, Eq. \ref{eq:slipshearlift-f}) . Present model predictions (\protect\blackline, Eq. \ref{eq:newmodel_test}).}
\label{fig:lift_constantepsi_out}
\end{figure}

In order to better understand the outer-region behaviour, in figure \ref{fig:lift_constantepsi_out} we plot the computed lift force coefficients against the separation distance normalised using the Saffman's length scale ($l^{\ast}/L_{\text{G}}^{\ast}$)
. The numerical results are compared against two existing outer-region-based correlations, namely $C_\text{L,23}^\text{wb,out}$ given via Eq. (\ref{eq:slipshearlift-outercl2cl3}) and $C_\text{L,2}^\text{wb,out}$ given via Eq. (\ref{eq:slipshearlift-f}). However, while $C_\text{L,23}^\text{wb,out}$ is valid for both small and large $\epsilon$ values, $C_\text{L,2}^\text{wb,out}$ is only valid when shear dominates ($\epsilon > 1$). In both cases $J(\epsilon)$ is evaluated using \citet{Legendre98}'s correlation (Eq.  \ref{Leg}) and $f(\epsilon,l/L_\text{G})$ is evaluated using \citet{TakemuraMagnaudet1}'s correlation, Eq. (\ref{f_TM}). Eq. (\ref{TM2004}) is used to evaluate $C_\text{L,3}^{\text{wb,out}}$, present in the expression for $C_\text{L,23}^\text{wb,out}$. As $(l^{\ast}/L_\text{G}^{\ast})$ increases, the theoretical lift force results for positive and negative $\gamma^{\ast} (=\gamma/u_\text{slip})$, obtained by varying only the slip direction, asymptote to the negative and positive unbounded lift forces, respectively. Although the theoretical model given by Eq. (\ref{eq:slipshearlift-outercl2cl3}) fails to capture the numerical lift variation in the transition region ($l^{\ast}\sim L_\text{G}^{\ast}$) when $|\gamma^{\ast}| \geq 10$, the same model predicts the numerical data reasonably well when $|\gamma^{\ast}| \leq 1$. On the other hand, the predictions of Eq. (\ref{eq:slipshearlift-f}) coincide with the numerical data only when $|\gamma^{\ast}| \sim 1$ ($Re_\gamma \sim Re_\text{slip}$) (see figure \ref{fig:lift_constantepsi_positive_out}).

\begin{figure}
    \begin{subfigure}{\linewidth}
    \centerline{\includegraphics[width=0.85\textwidth]{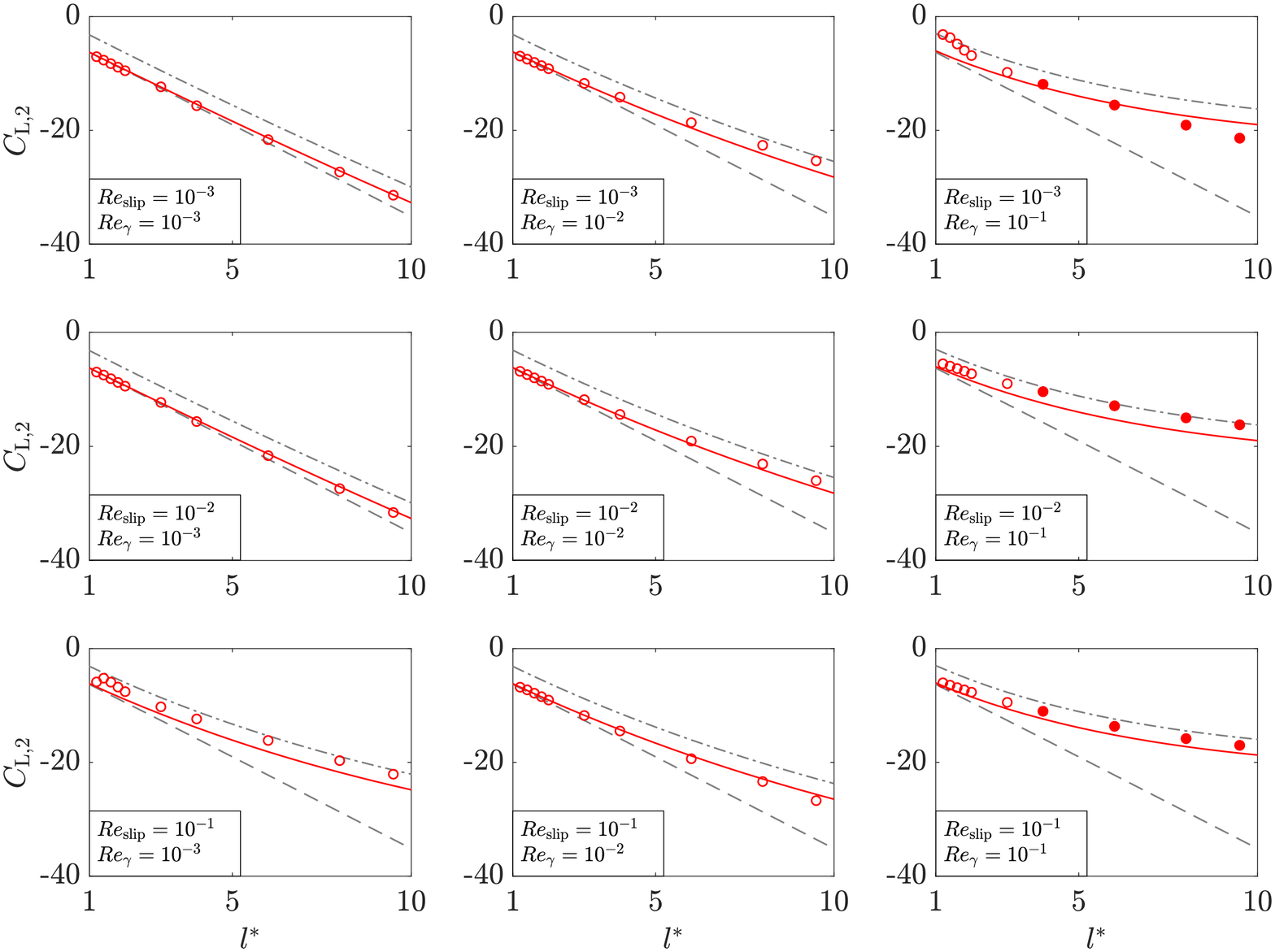}}
    \caption{Leading particle in a positive shear field (${\gamma}^{\ast} > 0$)}
    \label{fig:lift_cl2_only_positive}
    \end{subfigure}
    \begin{subfigure}{\linewidth}
    \medskip
    \centerline{\includegraphics[width=0.85\textwidth]{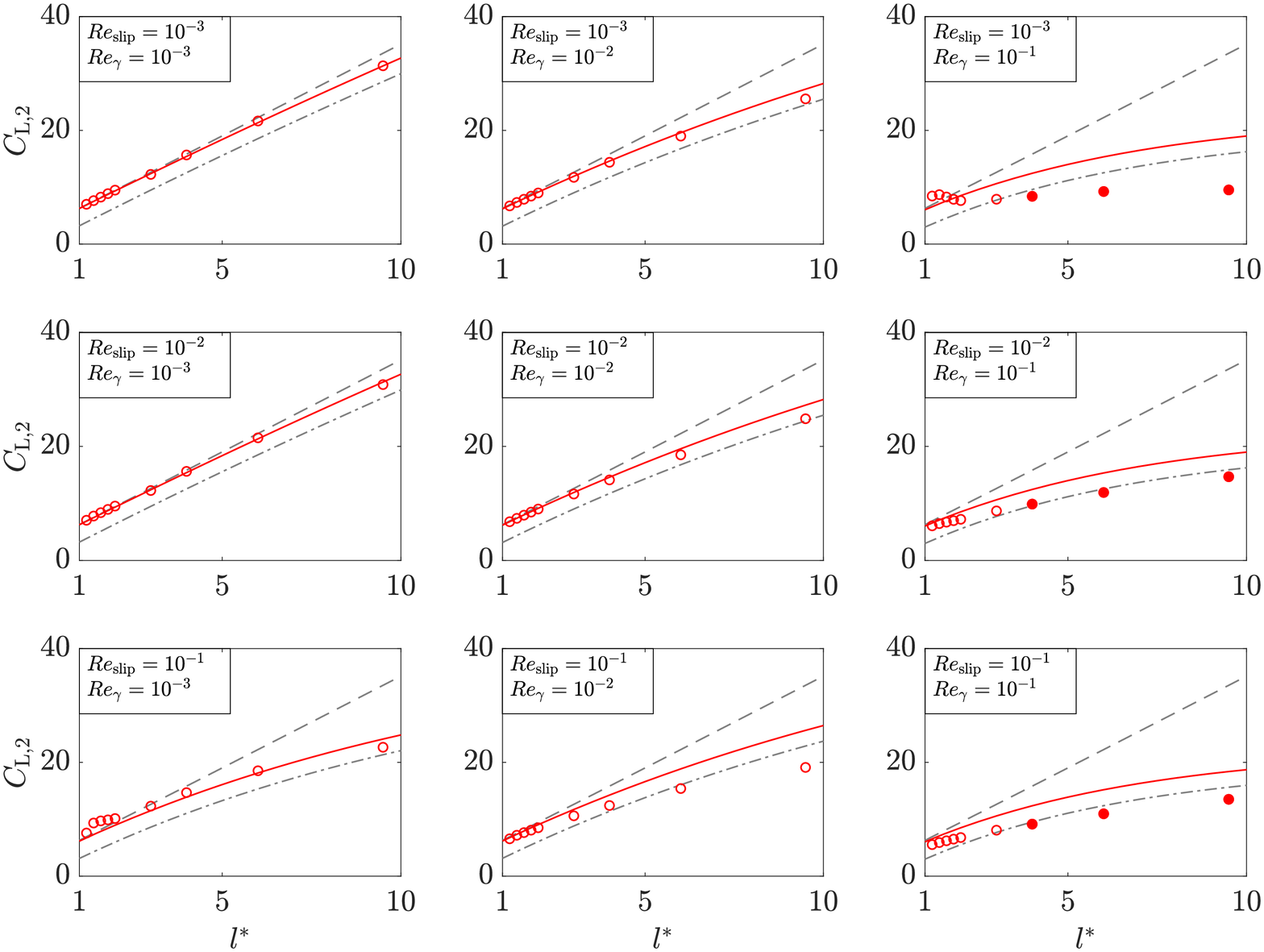}}
    \caption{Lagging particle in a positive shear field (${\gamma}^{\ast} < 0$)}
    \label{fig:lift_cl2_only_negative}
    \end{subfigure}
\caption{Lift force coefficient $C_{\text{L,2}}$ of a freely-rotating particle. $Re_\gamma$ and $Re_\text{slip}$ increase in the order of $10^{-3}, 10^{-2}, 10^{-1}$ from left to right and top to bottom respectively. Simulations: inner-region (\protect\hollowredcircle) and outer-region ($\textcolor{red}{\bullet}$) for each subplot. Analytical predictions: inner-region $C_{\text{L,2}}$ model of \citet{CherukatMcLaughlinCorrection} (\protect\Graydashedline, Eq. \ref{CM2}), outer-region model of \citet{TakemuraMagnaudet1} (\protect\Graydashdottedline, Eq. \ref{eq:slipshearlift-f}). Present model prediction (\protect\redline, Eq. \ref{eq:cl2_in-out_free})}
\label{fig:lift_cl2only}
\end{figure}

No existing models capture the force variation in both the inner and outer-regions successfully for all of the Reynolds numbers considered here. Recalling our definition for the net lift force given by Eq. (\ref{eq:jfm2:netlift_allregion_0}), since the coefficients $C_\text{L,1}$ and $C_\text{L,3}$ capture the lift contributions due to finite shear and finite slip conditions in the limits of $\gamma^{\ast} \rightarrow \infty$ and $\gamma^{\ast} = 0$, respectively, the remaining coefficient $C_\text{L,2}$ is found by subtracting the force contributions due to $C_\text{L,1}$ and $C_\text{L,3}$, from the present numerical results (see figure \ref{fig:lift_cl2only}). Here, $C_\text{L,1}$ and $C_\text{L,3}$ are evaluated using Eq. (3.3) in \citet{EkanayakeJFM1} and Eq. (\ref{eq:lift0shear_inout}) in the present study respectively.

\begin{figure}
    \begin{subfigure}{\linewidth}
    \centerline{\includegraphics[width=0.8\textwidth]{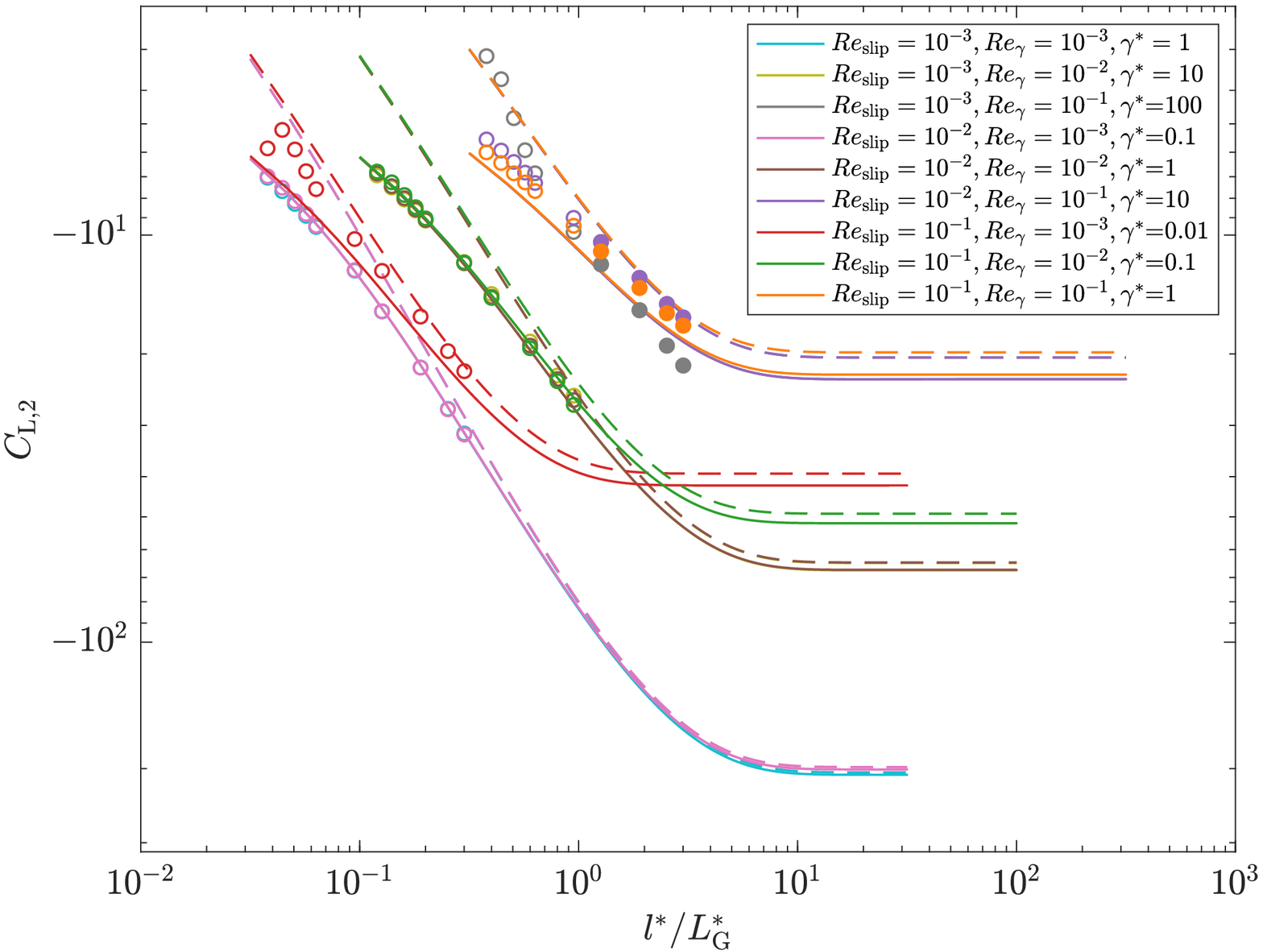}}
    \caption{Leading particle in a positive shear field (${\gamma}^{\ast} > 0$)}
    \label{fig:lift_cl2only_saffman_positive}
    \end{subfigure}
    \begin{subfigure}{\linewidth}
    \medskip
    \centerline{\includegraphics[width=0.8\textwidth]{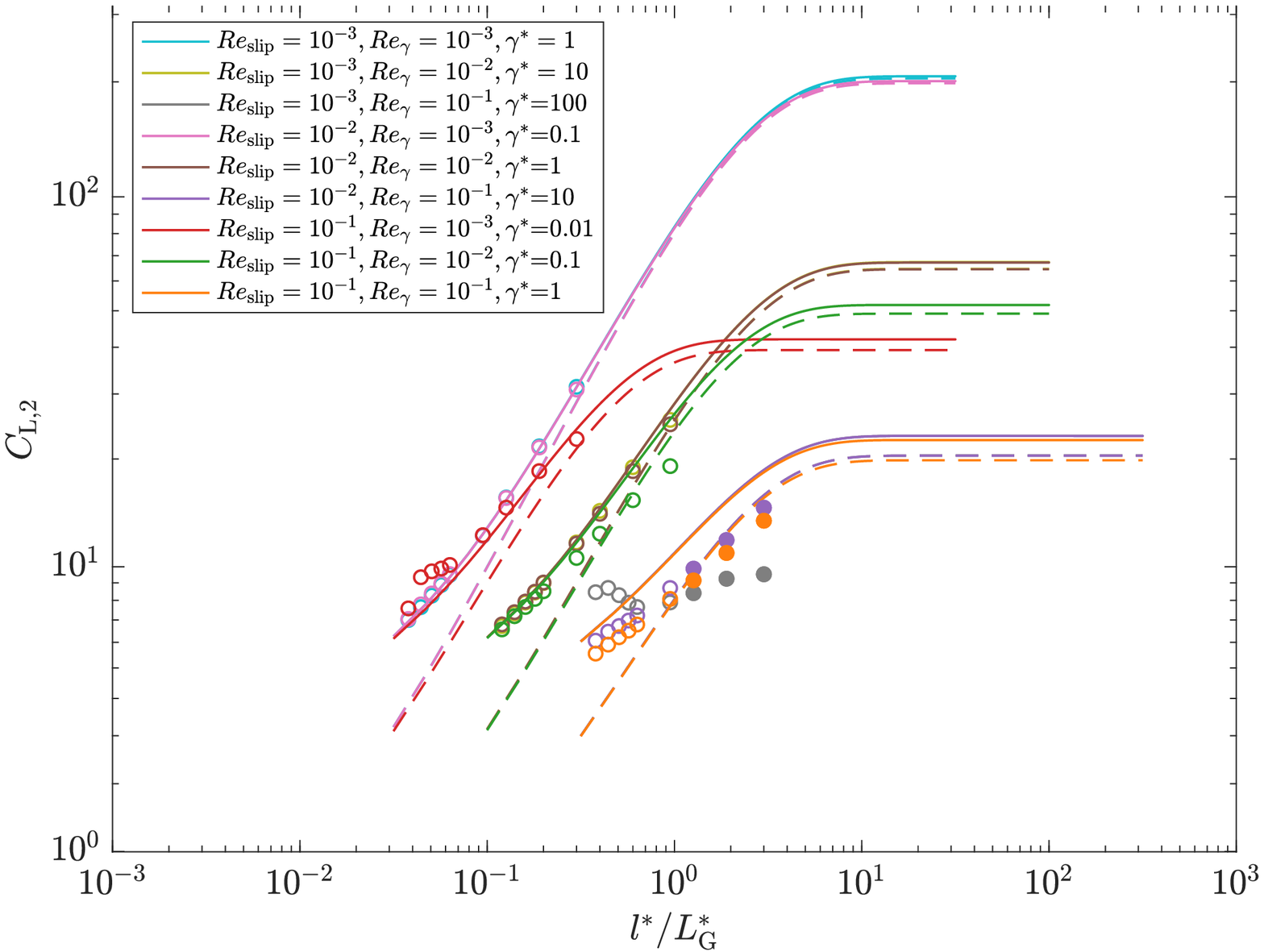}}
    \caption{Lagging particle in a positive shear field (${\gamma}^{\ast} < 0$)}
    \label{fig:lift_cl2only_saffman_negative}
    \end{subfigure}
\caption{Lift force coefficient $C_{\text{L,2}}$ of a freely-rotating particle as a function of separation distance non-dimensionalised by Saffman length scale ($l^{\ast}/{L_{\text{G}}}^{\ast}$). Simulations: inner-region (\protect\hollowcircle) and outer-region ($\bullet$). Analytical outer-region correlation of \citet{TakemuraMagnaudet1} (\protect\blackdashedline, Eq. \ref{eq:slipshearlift-f}). Present model predictions (\protect\blackline, Eq. \ref{eq:cl2_in-out}).}
\label{fig:lift_cl2only_saffman}
\end{figure}

We define a new correlation for $C_{\text{L,2}}$ in the following manner. To capture the variation of the remaining force contributions and the inner and outer-region transition behaviour, the outer-region-based, wall-bounded lift coefficient $C_\text{L,2}^\text{wb,out}$ given by Eq. (\ref{eq:slipshearlift-f}) is substituted into the lowest order term of the \citet{CherukatMcLaughlinCorrection}'s inner-region slip-shear lift $C_\text{L,2}^\text{wb,in}$ correlation given by Eq. (\ref{CM1}) for a non-rotating particle and Eq. (\ref{CM2}) for a freely-rotating particle. Noting that $C_\text{L,2}^\text{wb,out}$ in Eq. (\ref{eq:slipshearlift-f}) scales with $Re_\text{slip}^2$, the outer-region coefficient is divided by $|\gamma^{\ast}|$ to match the inner-region lift correlation scaling. The resulting slip-shear based net lift coefficient $C_\text{L,2}$, which is valid for all three regions (i.e., inner, outer and unbounded) is,
\begin{subequations}
\begin{align}
    C_\text{L,2}=-{C_\text{L,2}^\text{wb,out}}\frac{1}{\lvert{\gamma}^{\ast}\rvert} -1.1450-2.0840\bigg(\dfrac{1}{l^{\ast}}\bigg)+0.9059\bigg(\dfrac{1}{l^{\ast}}\bigg)^{2}\label{eq:cl2_in-out_nonrot}\\
\intertext{for a non-rotating particle and}
    C_\text{L,2}=-{C_\text{L,2}^\text{wb,out}}\frac{1}{\lvert{\gamma}^{\ast}\rvert} -2.6729-0.8373\bigg(\dfrac{1}{l^{\ast}}\bigg)+0.4683\bigg(\dfrac{1}{l^{\ast}}\bigg)^{2}
    \label{eq:cl2_in-out_free}
\end{align}
\label{eq:cl2_in-out}%
\end{subequations}
for a freely-rotating particle. The new $C_\text{L,2}$ correlation proposed for the freely rotating particle is plotted as a function of $l^{\ast}$ in figure \ref{fig:lift_cl2only} and as a function of normalised Saffman's length $l^{\ast}/L_\text{G}^{\ast}$ in figure \ref{fig:lift_cl2only_saffman}.
Given that the new correlation for $C_\text{L,2}$ (Eq. \ref{eq:cl2_in-out}) captures the variation of the remaining lift force contributions reasonably well for most of the slip and shear Reynolds numbers considered, we substitute this force model into the main net lift force correlation. The performance of the net lift correlation (Eq. \ref{eq:jfm2:netlift_allregion_0}) is then examined for a freely-rotating particle by plotting the force (${F}_\text{L}^{\ast}$) predictions in figure \ref{fig:lift_constantepsi}. The overall lift coefficient (${C}_\text{L}$), obtained by normalising the net lift force by the slip Reynolds number,
\begin{align}
    {C}_\text{L} = \frac{{F}_\text{L}^{\ast}}{Re_{\text{slip}}^2}  = {\gamma^{\ast}}^2 {C}_\text{L,1} + \gamma^{\ast}{C}_\text{L,2} + {C}_\text{L,3}
    \label{eq:newmodel_test}
\end{align}
is shown in figure \ref{fig:lift_constantepsi_out}. Here again, the lift coefficients, $C_\text{L,1}$ and $C_\text{L,3}$ are evaluated using Eq. (4.1) from our previous study \citep{EkanayakeJFM1} and (Eq. \ref{eq:lift0shear_0toque_inout}) from \S\ref{sec:jfm2:uniformflow}. For reference a summary of all the lift correlations used in the net lift force calculations are provided in Appendix \ref{appA}.

As shown in both figures \ref{fig:lift_constantepsi} and \ref{fig:lift_constantepsi_out}, the new model captures the inner-outer transition behaviour of the computed lift results well. Referring to figure \ref{fig:lift_constantepsi}, for low slip and shear values ($Re_\gamma$ and $Re_\text{slip} \lesssim 10^{-2}$), the model performs well for both positive and negative $\gamma^{\ast}$. For large slip values and for small shear rates (i.e., $Re_\text{slip} = 10^{-1}$ and $Re_\gamma = 10^{-3}$), the model slightly overestimates (underestimates) the simulated results with a maximum deviation of $4.96\%$ $(3.67\%)$ for $\gamma^{\ast} > 0$ ($\gamma^{\ast} < 0$) when the particle is furthest from the wall. With increasing shear rate (i.e., $Re_\gamma \sim 10^{-1}$) for the same large slip values, this deviation rapidly reduces for ${\gamma}^{\ast} > 0$, but increases to $24.03\%$ for ${\gamma}^{\ast} < 0$. 

As well as the force magnitudes, the change of the lift force direction is more accurately predicted (i.e., ${Re_\gamma} = 10^{-1}$) using the new correlation than any other available model. Note that a positive lift ($F_\text{L}^{\ast}>0$) and a negative lift ($F_\text{L}^{\ast} <0$) represent a force directed away from and a force acting towards the wall, respectively. In figure \ref{fig:lift_constantepsi_positive}, the lift forces computed for ${\gamma}^{\ast} > 0$ indicate a change of the lift force direction and a decrease in the force with increasing separation distance. However, for the same $Re_\gamma$ and $Re_\text{slip}$ values, the numerical data given in figure \ref{fig:lift_constantepsi_negative} for ${\gamma}^{\ast} < 0$, only indicate positive lift forces for the selected range of separation distances, and generally an increase in the lift force with increasing separation distance.

The force variations shown in figures \ref{fig:lift_constantepsi} and \ref{fig:lift_constantepsi_out} can be explained by examining the behaviour of the three theoretical lift coefficients ($C_\text{L,1}, C_\text{L,2}$ and $C_\text{L,3}$) separately. The coefficients $C_\text{L,1}$ and $C_\text{L,3}$, responsible for the lift due to pure shear and pure slip respectively, always remain positive irrespective of the direction of both slip and shear. Thus, the lift forces due to these two coefficients will always act to push a particle away from the wall. However, with increasing separation distance, the values of these two lift coefficients rapidly reduce to zero \citep{EkanayakeJFM1,VasseurCox2}. Therefore $C_\text{L,1}$ and $C_\text{L,3}$ are only important close to the wall. The remaining slip-shear lift coefficient, $C_\text{L,2}$, behaves differently. Unlike the other two coefficients, $C_\text{L,2}$ is sensitive to the direction of slip and shear (determined by sgn(${\gamma}^{\ast}$)), and also has a finite negative or positive value in the unbounded limit. Hence, at large separation distances, the net lift force mainly depends on the $C_\text{L,2}$ coefficient and its corresponding sign. However near a wall, the net lift force magnitude and the direction strongly depend upon the inner-region contributions of all three coefficients. 


Including all lift contributions covering the inner and outer-regions means that the present lift model predicts the correct lift coefficient variation with $Re_\gamma, Re_\text{slip}$ and $l^{\ast}$, over the wide range of parameters considered in this study (Figs \ref{fig:lift_constantepsi} \& \ref{fig:lift_constantepsi_out}).

\subsubsection{Drag force}
\begin{figure}
    \begin{subfigure}{\linewidth}
    \centerline{\includegraphics[width=0.8\textwidth]{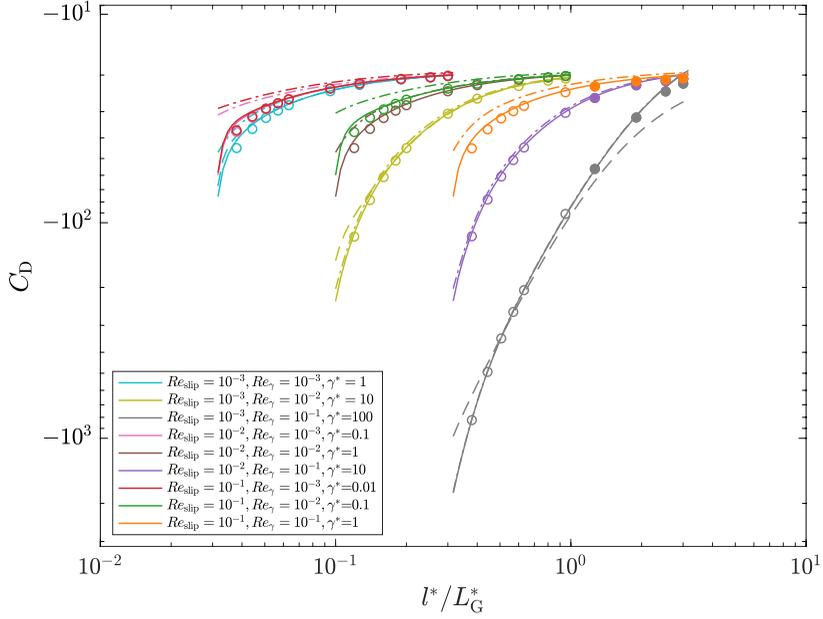}}
    \caption{Leading particle in a positive shear field (${\gamma}^{\ast} > 0$)}
    \label{fig:drag_constantepsi_positive_out}
    \end{subfigure}
    \begin{subfigure}{\linewidth}
    \medskip
    \centerline{\includegraphics[width=0.8\textwidth]{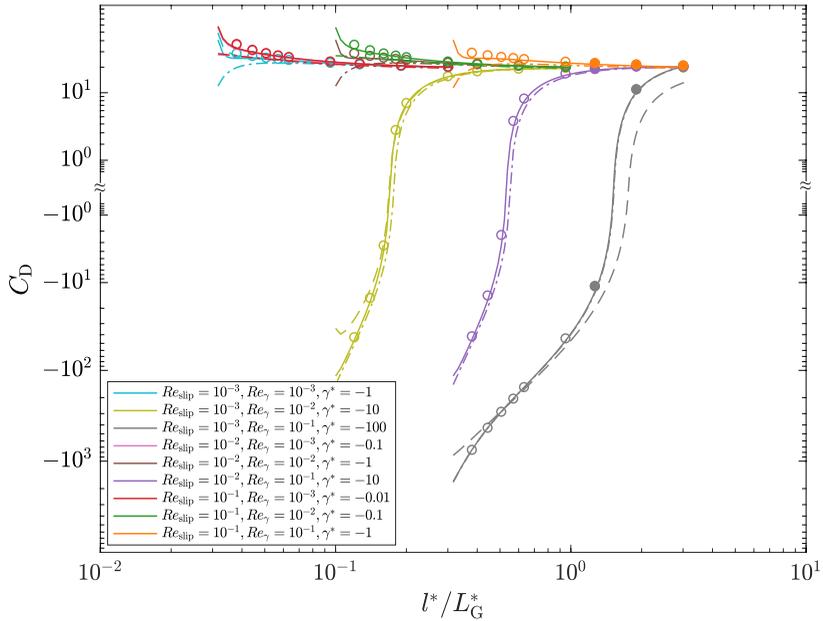}}
    \caption{Lagging particle in a positive shear field (${\gamma}^{\ast} < 0$)}
    \label{fig:drag_constantepsi_negative_out}
    \end{subfigure}
\caption{Drag force coefficient ($C_{\text{D}}$) of a freely-rotating particle as a function of separation distance non-dimensionalised by Saffman length scale ($l^{\ast}/{L_{\text{G}}}^{\ast}$). Simulations: inner-region (\protect\hollowcircle) and outer-region ($\bullet$). inner-region analytical predictions when $C_\text{D,1}$ is evaluated by Eq. (\ref{eq:drag_cd1_nilanka}) (\protect\blackline) and by Eq. ( \ref{eq:drag_cd1_magneduet}) (\protect\blackdashedline). For both cases $C_\text{D,2}$ is evaluated by Eq. (\ref{eq:drag_wb_faxen2}). Outer-region analytical prediction when $C_\text{D,2}$ and $C_\text{D,1}$ are evaluated by Eq. (\ref{eq:drag_wb_takemura_outer}) and Eq. (\ref{eq:drag_cd1_nilanka}) respectively (\protect\blackdashdottedline).}
\label{fig:drag_constantepsi_out}
\end{figure}

The drag force on a freely-rotating spherical particle moving parallel to a wall in a linear shear flow is analysed in this section. Here, the drag force normalised by the slip Reynolds number
\begin{align}
    {C}_\text{D} = \frac{{F}_\text{D}^{\ast}}{Re_\text{slip}} = -\gamma^{\ast}C_\text{D,1}-C_\text{D,2}
    \label{eq:net_drag}    
\end{align}
is used to define the net drag coefficients. Results are presented for positive and negative slip velocities in a positive shear field, noting that results for a negative shear field are identical to the presented positive shear field results with the sign of the slip velocity swapped. Figure \ref{fig:drag_constantepsi_out} shows the variation in the net drag coefficient $C_\text{D}$ for slip and shear Reynolds numbers in the range $10^{-3} - 10^{-1}$ as a function of wall separation distance normalised by Saffman's length. A positive slip in the presence of the wall produces a force on a leading particle that is in the opposite direction to the flow (Figure \ref{fig:drag_constantepsi_positive_out}). The largest negative values for $C_\text{D}$ are obtained when the particle is close to the wall, reducing to the unbounded Stokes drag result with increasing wall distance. Both positive and negative values for $C_\text{D}$ are reported for a lagging particle close to the wall (Figure \ref{fig:drag_constantepsi_negative_out}), with the direction of the force depending on both the $C_\text{D,1}$ and $C_\text{D,2}$ magnitudes. Although a lagging particle results in a positive slip-drag force contribution, the positive shear produces a force in the opposite direction to the flow. As a result negative net drag forces are obtained near the wall at high $\gamma^{\ast}$.

The simulated drag coefficient results are also compared against the inner-region and outer-region-based theoretical drag correlations at low $Re_{\gamma} \ll 1$. Two correlations, Eqs. (\ref{eq:drag_cd1_nilanka} \& \ref{eq:drag_cd1_magneduet}), for $C_\text{D,1}$ are first tested while using an inner-region-based Faxen drag coefficient for $C_\text{D,2}$ (Eq. \ref{eq:drag_wb_faxen2}). As shown in figure \ref{fig:drag_constantepsi_out}, Eq. (\ref{eq:drag_cd1_nilanka}) for $C_\text{D,1}$ performs better than Eq. (\ref{eq:drag_cd1_magneduet}) when predicting the lift results for high $\gamma^{\ast}$. However, no significant difference between these two models is observed for small $\gamma^{\ast}$. The outer-region-based drag model of \citet{Takemura} (Eq. \ref{eq:drag_wb_takemura_outer}) is also tested for $C_\text{D,2}$ in combination with Eq. (\ref{eq:drag_cd1_nilanka}) for $C_\text{D,1}$. Unsurprisingly, this model fails to capture the inner-region behaviour in either slip direction, particularly closer to the wall, although the model predictions agree reasonably well with other theoretical models for large $l^{\ast}/{L_{\text{G}}}^{\ast}$ values. In summary, for the considered range of slip and shear Reynolds numbers, the most accurate drag predictions are obtained using the \citet{EkanayakeJFM1} model (Eq. \ref{eq:drag_cd1_nilanka}) for $C_\text{D,1}$ and the \citet{Faxen22} model (Eq. \ref{eq:drag_wb_faxen2}) for $C_\text{D,2}$, respectively.

\section{Application of the combined model: Buoyant and Non-buoyant particles}\label{sec:jfm2:Application}
In this section, we examine the movement of both force-free (neutrally-buoyant) and buoyant particles in a linear shear flow using the new correlations. We also examine the movement of buoyant particles in a quiescent flow. To validate the force-free results, we compare against the numerical results of \citet{EkanayakeJFM1}. To validate the buoyant results we use results of two previous experimental studies conducted by \citet{Takemura} and \citet{TakemuraMagnaudet2} for rigid spherical particles at low but finite slip Reynolds numbers ($0.05 < Re_\text{slip} < 2.5$). Given the scope of the current study, we only use the experimental data where $Re_\text{slip} < 1$.

\subsection{Force-free particle in a linear shear flow}

\begin{figure}
\centerline{\includegraphics[width=0.8\textwidth]{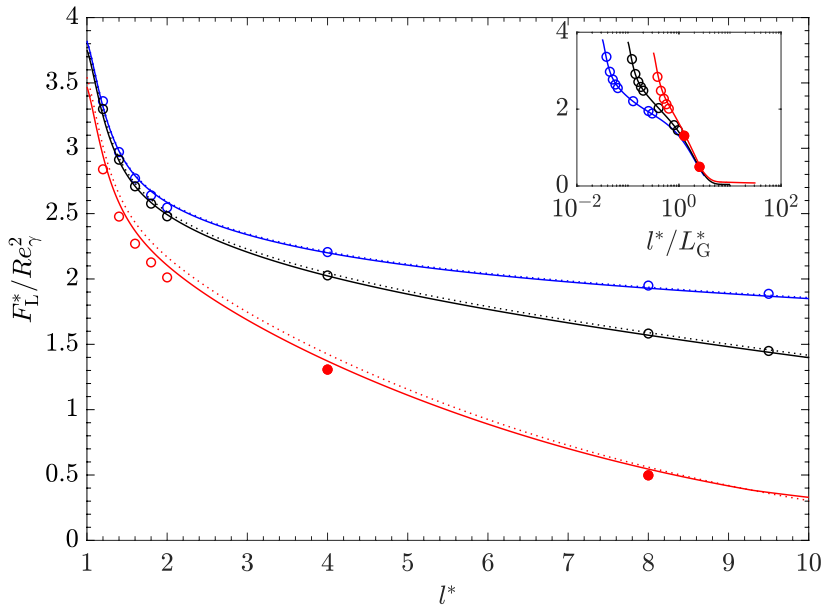}}
\caption{Lift force of a force-free particle translating in a linear shear flow near a wall when ${F_\text{L,tot}^{\prime}}$ is evaluated using Eq. (\ref{eq:net_lift_neutrallybouyant}) for $Re_\gamma=10^{-3}$ (\protect\blueline), $Re_\gamma=10^{-2}$ (\protect\blackline), $Re_\gamma=10^{-1}$ (\protect\redline); Numerical results  \citep{EkanayakeJFM1} (coloured hollow circles). $C_\mathrm{L,2}$ and $C_\mathrm{L,3}$ in Eq. (\ref{eq:net_lift_neutrallybouyant}) are evaluated using inner-region-based \citet{CherukatMcLaughlinCorrection}'s models (dotted lines).}
\label{fig:jfm2:nb_lift}
\end{figure}

In this section, we re-examine the movement of the force-free particle which has been previously discussed in \citet{EkanayakeJFM1}. However, we now employ the proposed $C_\mathrm{L,2}$ and $C_\mathrm{L,3}$ coefficients in Eq. (\ref{eq:jfm2:netlift_allregion_0}), which span across all three regions. In this section, the net lift force coefficient obtained by normalising the net lift force (Eq. \ref{eq:jfm2:netlift_allregion_0}) using the shear Reynolds number, 
\begin{align}
    {C}_\text{L} = \frac{{F}_\text{L}^{\ast}}{Re_{\gamma}^2}  =  {C}_\text{L,1} + \frac{1}{\gamma^{\ast}}{C}_\text{L,2} + \frac{1}{{\gamma^{\ast}}^2}{C}_\text{L,3}
    \label{eq:net_lift_neutrallybouyant}
\end{align}
is used to present lift results. The procedure followed to calculate the slip velocity of the force-free particle is same as in \citet{EkanayakeJFM1}. Note that the $C_\text{D,2}$ correlation used in our previous study is further validated for linear shear flows in the present study (see \S \ref{sec:jfm2:constantepsi}). The calculated slip velocities are applied to Eq. (\ref{eq:net_lift_neutrallybouyant}) together with the new ${C}_\text{L,2}$ and ${C}_\text{L,3}$ lift coefficients. The lift results are plotted in figure \ref{fig:jfm2:nb_lift}, and compared against the direct numerical numerical results. The previous estimations which used the inner-region-based correlations for $C_\mathrm{L,2}$ and $C_\mathrm{L,3}$ are also plotted in the same figure.

The predictions from Eq. (\ref{eq:net_lift_neutrallybouyant}) agree reasonably well with the numerical results. Near the wall ($l^{\ast} < 2$), the present lift model predictions are more accurate than the previous estimations, particularly for the highest shear Reynolds number. Also, the inset of the figure illustrates that the net lift of a force-free particle rapidly reduces to zero as the separation distance increases.

\subsection{Buoyant particle in a quiescent fluid}
\begin{figure}
    \begin{subfigure}{\linewidth}
    \centerline{\includegraphics[width=0.8\textwidth]{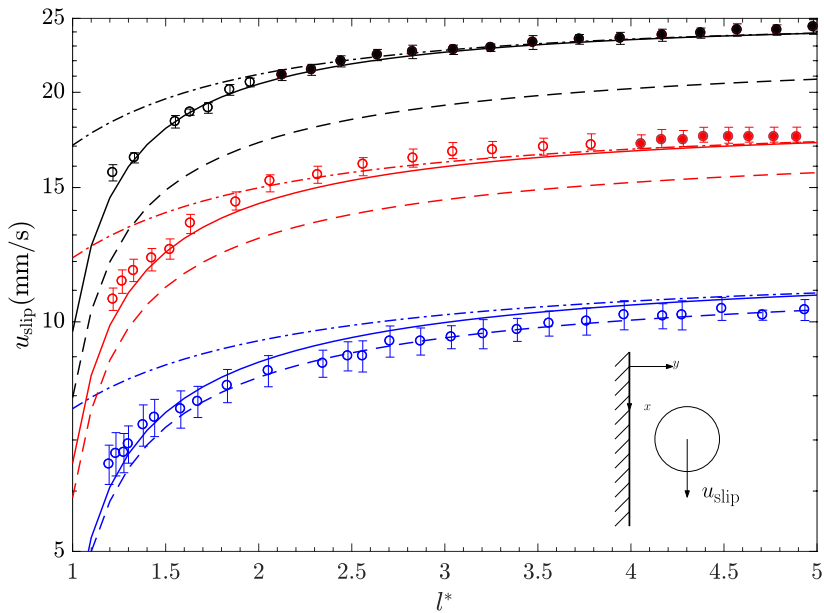}}
    \caption{Slip velocity when $C_\text{D,2}^\text{wb}$ is evaluated using, Eq. (\ref{eq:drag_wb_faxen2}) (dashed line), Eq. (\ref{eq:drag_wb_takemura_outer}) (dashed and dotted line) and Eq. (\ref{eq:drag_wb_takemura_innerouter}) (solid line) and $C_\text{D,2}^\text{ub}$ is evaluated by Eq. (\ref{IC_Clift}).}
    \label{fig:slip_takemura0shear}
    \end{subfigure}
    \begin{subfigure}{\linewidth}
    \medskip
    \centerline{\includegraphics[width=0.8\textwidth]{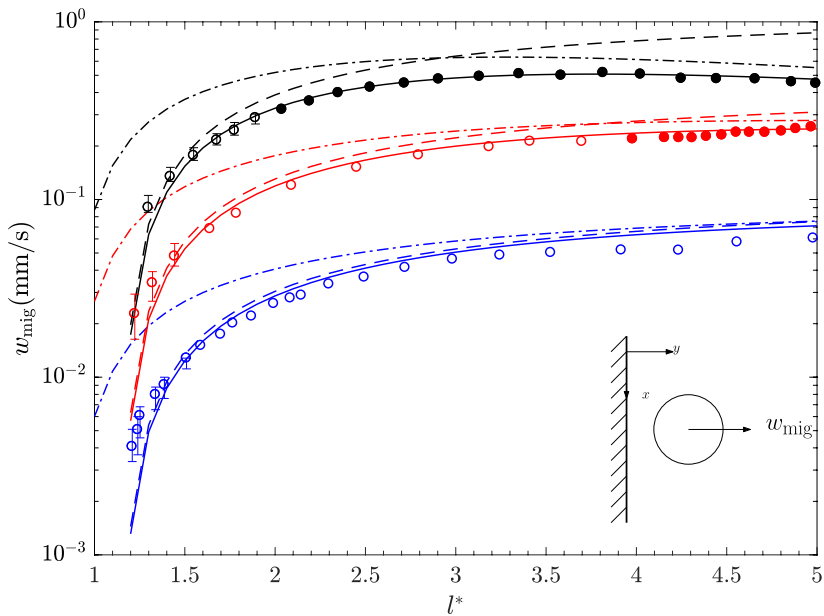}}
    \caption{Migration velocity when $C_\text{L,3}$ is evaluated by present model Eq. (\ref{eq:lift0shear_0toque_inout}) solid line) and by using inner-region correlation Eq. (\ref{MD1}) (dashed line) and outer-region correlation Eq. (\ref{TM2004}) (dashed and dotted line) as given in \citet{Takemura}.}
    \label{fig:migration_takemura0shear}
    \end{subfigure}
\caption{Analysis of a sedimenting particle in a quiescent single wall-bounded flow. Experiments: $Re_{\text{slip},\infty}=0.1005$ (\protect\hollowbluecircle), $Re_{\text{slip},\infty}=0.255$ (\protect\hollowredcircle), $Re_{\text{slip},\infty}=0.5$ (\protect\hollowcircle). Outer-region data are indicated using solid circles, and inner-region data by hollow circles.}
\label{fig:takemura0shear}
\end{figure}

The migration of a small particle falling near a wall in a quiescent flow is analysed using the lift correlations proposed in \S \ref{sec:jfm2:uniformflow}. For this, the slip velocities of the sedimenting particle are first calculated by balancing the buoyancy force, $F_\text{G} =4/3\pi a^3(\rho_\text{s}-\rho_\text{f})g$ with the wall-bounded fluid drag force, $F_\text{D}$ \citep{Takemura}. Here $\rho_\text{s}$ and $\rho_\text{f}$ are the solid and fluid densities respectively and $g$ is the gravity. In the original experimental study, the unbounded slip Reynolds numbers, $Re_{\text{slip},\infty} (=a u_{\text{slip},\infty}/\nu)$, were calculated by balancing $F_\text{G}$ with the unbounded fluid drag force, $F_{\text{D},\infty} (=6\pi \mu a u_{\text{slip},\infty}$). The local slip velocity that varies with the separation distance can hence can be written in terms of this unbounded slip velocity as:
\begin{equation}
u_\text{slip}=-6\pi\frac{\nu}{a}\Bigg(\frac{Re_{\text{slip},\infty}}{C_\text{D}}\Bigg)
\label{eq:slip_quiescent}
\end{equation}
Figure \ref{fig:slip_takemura0shear} shows the calculated and measured slip values \citep{Takemura} as a function of $l^{\ast}$ for three different values of $Re_{\text{slip},\infty}$. For quiescent flows, the net drag coefficient, $C_\text{D}$ given in Eq. (\ref{eq:slip_quiescent}) reduces to $-C_\text{D,2}$ according to Eq. (\ref{eq:net_drag}). For comparison, three correlations are used to evaluate the wall-bounded drag coefficient, $C_\text{D,2}^\text{wb}$, that capture the inner-region, outer-region and inner-outer-region transition behaviours. The inertial correction for $C_\text{D,2}^\text{ub}$ is evaluated by Eq. (\ref{IC_Clift}). Note that when plotting these figures, for the lowest slip Reynolds number, the $Re_{\text{slip},\infty} = 0.09$ value quoted in \citet{Takemura} had to be adjusted to $Re_{\text{slip},\infty} = 0.1005$ to get the correspondence given in \citet{Takemura}, based on the kinetic viscosity and density values combinations given in the original paper.

The overall combined inner-outer correlation predicts the $C_\text{D}$ well over all parameter ranges considered. However, the results for $Re_{\text{slip},\infty} = 0.1005$ in  figure \ref{fig:slip_takemura0shear} indicate that the Faxen's inner-region drag correlation given by Eq. (\ref{eq:drag_wb_faxen2}) does a better job at predicting the slip velocity, noting the modification we used on the reported $Re_{\text{slip},\infty}$. For the two larger $Re_{\text{slip},\infty}$, Eq. (\ref{eq:drag_wb_takemura_innerouter}) captures both inner-region and outer-region data points reasonably well, while for these particular cases Eqs. (\ref{eq:drag_wb_faxen2}) and (\ref{eq:drag_wb_takemura_outer}), deviate significantly from one another.

The derived slip velocities from Eq. (\ref{eq:slip_quiescent}) are then combined with the new lift correlation given for a freely-rotating particle in a quiescent flow (Eq. \ref{eq:newmodel_test}). Since the measured quantity was actually the dimensional transverse migration velocity of the particle relative to the wall ($w_\text{mig}$), a force balance normal to the wall is performed \citep{Takemura}. The forces considered here are the fluid drag force normal to the wall ($F_{\text{D}\perp} = \mu a w_\text{mig}C_{\text{D}\perp}$) and the particle lift force ($F_\text{L}$). 
\begin{align}
 w_\text{mig}=u_\text{slip}Re_\text{slip}\frac{C_\text{L}}{C_{\text{D}\perp}}
\label{eq:migrationvelocity_quiescent}
\end{align}
Here $C_{\text{D}\perp}$ is the wall-bounded drag coefficient of a particle translating normal to a wall. Similar to $C_{\text{D,2}}$, Faxen \citep{Happel} provided an analytical inner-region correlation for $C_{\text{D}\perp}$ while \citet{Takemura} provided an empirical fit to the outer-region correlations for $Re_\text{slip} \ll 1$. Since the reported $w_\text{mig}$ are small compared to $u_\text{slip}$, the inner-region correlation proposed by Faxen \citep{Happel} is used to calculate $C_{\text{D}\perp}$ in Eq. (\ref{eq:migrationvelocity_quiescent}):
\begin{align}
{C_{\text{D}\perp}}=\frac{6\pi}{1-\frac{9}{8}\bigg(\frac{1}{l^{\ast}}\bigg)+\frac{1}{2}\bigg(\frac{1}{l^{\ast}}\bigg)^{3}-\frac{135}{256}\bigg(\frac{1}{l^{\ast}}\bigg)^{4}-\frac{1}{8}\bigg(\frac{1}{l^{\ast}}\bigg)^{5}}
\label{eq:CDperp}
\end{align}
Note that the lift coefficient $C_\text{L}$ in Eq. (\ref{eq:migrationvelocity_quiescent}) reduces to $C_\text{L,3}$ due to the quiescent flow condition and the corresponding coefficients are evaluated using the new correlation Eq. (\ref{eq:lift0shear_0toque_inout}). 

The calculated migration velocity values are shown for different $Re_{\text{slip},\infty}$ in figure \ref{fig:migration_takemura0shear} and compared against the previous inner and outer-region lift models suggested in the original paper \citep{Takemura}. The theoretical and experimental migration velocity values show a strong dependence on both $Re_{\text{slip},\infty}$ and $l^{\ast}$. The analytical predictions using the new inner-outer based lift correlation (Eq. \ref{eq:lift0shear_0toque_inout}) agree reasonably well with the experimental results for all three $Re_{\text{slip},\infty}$ numbers. Although the pure inner-region-based predictions are fairly accurate for low slip Reynolds number (i.e, $Re_{\text{slip},\infty} = 0.1005$), the predictions  for larger $Re_{\text{slip},\infty}$ values deviate from the experimental results at larger separation distances. The outer-region-based migration velocity predictions are less accurate for all the examined cases. Interestingly, with increasing $l^{\ast}$, the experimental migration velocity results obtained for $Re_{\text{slip},\infty} = 0.5$ reach a maximum around $l^{\ast} \sim 2- 2.5$. For this particular slip Reynolds number, the transition from inner to outer-region also happens at $l^{\ast} \sim 2$, and the predictions based on the new lift correlation capture this transition behaviour more accurately than the other available models. Experimental values very close to the wall ($l^{\ast} \sim 1.2$) are slightly higher than the theoretical prediction. Nevertheless, notable measurement deviations were also reported closer to the wall, suggesting one potential cause for the discrepancies between experimental and theoretical values here.


\subsection{Buoyant particle in a linear shear flow}

\begin{figure}
    \begin{subfigure}{\linewidth}
    \centerline{\includegraphics[width=0.8\textwidth]{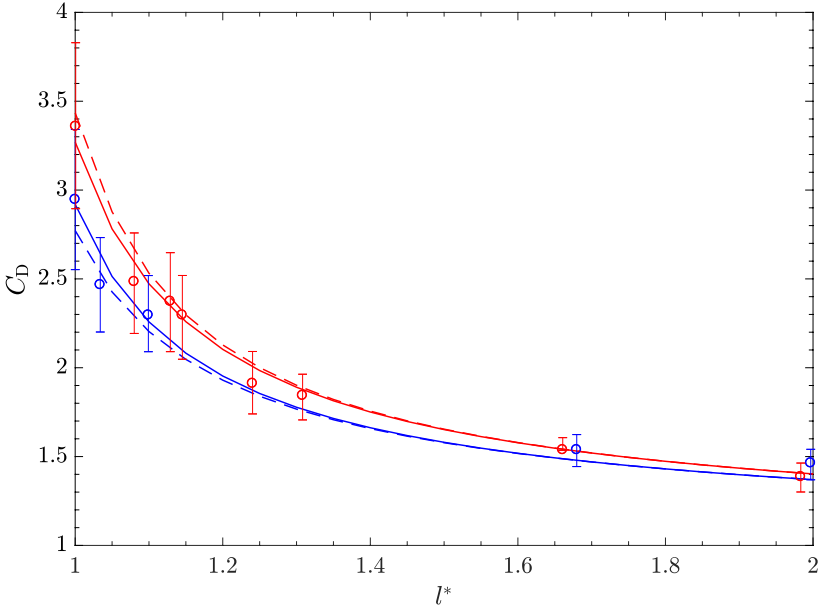}}
    \caption{Drag coefficient when $C_\text{D,1}$ is evaluated by Eq. (\ref{eq:drag_cd1_nilanka}) (solid) and Eq. (\ref{eq:drag_cd1_magneduet}) (dashed). For all cases $C_\text{D,2}$ is evaluated by Eqs. (\ref{eq:drag_wb_faxen2}, \ref{IC_Stokes}). Experiments: $Re_{\text{slip},\infty}=0.029, {\gamma}^{\ast}_\infty = -0.116$ (\protect\hollowbluecircle), $Re_{\text{slip}},\infty=0.029, {\gamma}^{\ast}_\infty = 0.116$ (\protect\hollowredcircle).}
    \label{fig:slip_takemuralinearshear}
    \end{subfigure}
    \begin{subfigure}{\linewidth}
    \medskip
    \centerline{\includegraphics[width=0.8\textwidth]{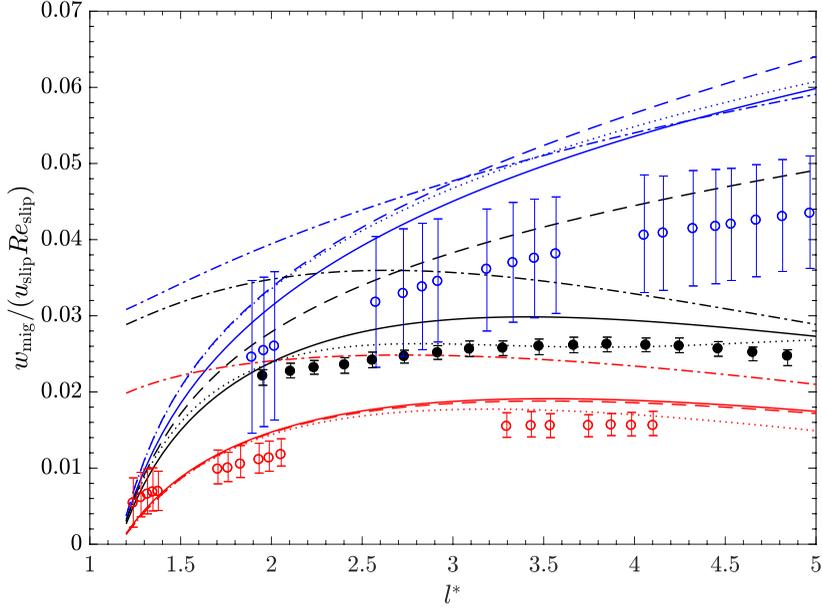}}
    \caption{Migration velocity when $C_\text{L}$ is evaluated by using proposed lift Eq. (\ref{eq:newmodel_test}) (solid), inner-region model Eq. (\ref{eq:liftinner}) using \citet{Magnaudet1} coefficients (dashed), outer-region model Eq. (\ref{eq:slipshearlift-outercl2cl3}) using \citet{TakemuraMagnaudet1} coefficients (dashed dotted) and empirical fit \citet{Takemura} (dotted). Experiments: $Re_{\text{slip},\infty}=0.1, {\gamma}^{\ast}_\infty = -0.061$ (\protect\hollowbluecircle), $Re_{\text{slip},\infty}=0.17, {\gamma}^{\ast}_\infty = 0.044$ (\protect\hollowredcircle), $Re_{\text{slip}},\infty=0.55, {\gamma}^{\ast}_\infty = -0.033$ (\protect\hollowcircle). Outer-region data are indicated using solid circles, and inner-region data by hollow circles.}
    \label{fig:migration_takemuralinearshear}
    \end{subfigure}
\caption{Analysis of a sedimenting particle in a linear single wall-bounded flow}
\label{fig:takemuralinearshear}
\end{figure}

In this section a small spherical particle falling in a linear shear flow near a wall is examined using the lift and drag correlations from \S \ref{sec:jfm2:constantepsi}. Both positive and negative shear flows are considered.  

First, the net drag coefficient for positive and negative shear rates of the same magnitude are analysed at small slip ($Re_{\text{slip}}=0.029$). The variations are compared against experimentally measured drag coefficient values \citep{TakemuraMagnaudet2}. Unlike the quiescent situation, here $C_\text{D}$ is a function of both $C_\text{D,1}$ and $C_\text{D,2}$. We compare two different correlations for $C_\text{D,1}$ (Eqs. \ref{eq:drag_cd1_magneduet} and \ref{eq:drag_cd1_nilanka}) and use the inner-region-based Faxen drag correlation (Eq. \ref{eq:drag_wb_faxen2}) for $C_\text{D,2}$. 

As expected, the net drag coefficient of a sedimenting particle increases when the particle moves closer to the wall. However, higher drag coefficients are observed for a sedimenting particle in a positive shear ($\gamma^{\ast} > 0$), when compared to the negative shear field ($\gamma^{\ast} < 0$). This variation decreases as separation distance increases since the effect of $C_\text{D,1}$ rapidly decays with $l^{\ast}$. A relatively small difference is observed when using Eq. (\ref{eq:drag_cd1_magneduet}) compared to Eq. (\ref{eq:drag_cd1_nilanka}) to calculate $C_\text{D,1}$, a result of shear being relatively low in this analysed system. 

Next the lift correlations derived for linear shear flow at $Re_{\text{slip},\infty}\leq\mathcal{O}(10^{-1})$ are used to predict migration velocities. Similar to the previous section, the slip velocity of the sedimenting particle is first calculated using Eq. (\ref{eq:slip_quiescent}). The values are then combined with the new lift correlation given for a freely-rotating particle to find the measured dimensionless transverse migration velocity of the particle relative to the fluid ($w_\text{mig}/(u_\text{slip} Re_\text{slip})$),
\begin{align}
 \frac{w_\text{mig}}{u_\text{slip}Re_\text{slip}} =\frac{C_\text{L}}{C_{\text{D}\perp}}
\label{eq:migrationvelocity_linear}
\end{align}
Here ${C_{\text{D}\perp}}$ is again evaluated using the Faxen inner-region expression Eq. (\ref{eq:CDperp}) and $C_\text{L}$ is evaluated using the new correlation given in Eq. (\ref{eq:newmodel_test}).

Figure \ref{fig:migration_takemuralinearshear} shows the dimensionless migration velocity as a function of normalised separation distance for three different $Re_{\text{slip},\infty}$, in which two cases are for negative shear rates. All the data reported for the lowest two $Re_{\text{slip},\infty}$ numbers are in the inner-region as the dimensionless Stokes length extends up to $l^{\ast} = 10$ for $Re_{\text{slip},\infty} = 0.1$ and $l^{\ast} = 5.9$ for $Re_{\text{slip},\infty} = 0.17$. For these two experiments, note that the Saffman length is much larger than the Stokes length since $\epsilon \ll 1$. For $Re_{\text{slip},\infty} = 0.55$, the inner-region shrinks as the non-dimensional Stokes length reduces to $l^{\ast} = 1.81$ while the non-dimensional Saffman length spans up to $l^{\ast} = 5.5$. Therefore all the experimental data for this case are in the outer-region according to the definition of the region boundary at $l^{\ast} = \text{min}(L_\text{S}^\ast,L_\text{G}^\ast)$. Based on experimental results, \citet{TakemuraMagnaudet1} suggested an empirical fit for the migration velocity by combining the lift and the wall normal drag coefficients. Although the migration velocity predictions of this fit closely follow the experimental results (i.e., dotted lines in figure \ref{fig:migration_takemuralinearshear}), this correlation does not provide information about individual forces acting on the particle, and hence this fit cannot be used to predict the particle behaviour when different hydrodynamic forces are acting on a particle simultaneously.

The quantitative agreement of the present model against the experimental data is actually better for the largest two $Re_{\text{slip},\infty}$ values, than for the $Re_{\text{slip},\infty} = 0.1$ case. However, as the migration velocity is very small for $Re_{\text{slip},\infty} = 0.1$, \citet{TakemuraMagnaudet1} suggested that the experimental measurements could be less reliable for the entire range of $l^{\ast}$ for this case. For $Re_{\text{slip},\infty} = 0.17$ and $0.55$, the proposed lift correlation captures both the inner and outer-region behaviour in both negative and positive shear environments reasonably well. Unsurprisingly, Eq. (\ref{eq:newmodel_test}) follows the inner-region model prediction of \citet{TakemuraMagnaudet1} for $Re_{\text{slip},\infty} = 0.17$ as all the experimental measurements are well within the inner-region. Interestingly, the calculated migration velocities for $Re_{\text{slip},\infty} = 0.55$, using the new lift correlation, capture the experimental outer-region behaviour fairly well in the region between the normalised Stokes and Saffman lengths ($1.82<l^{\ast}<5.5$). 


\section{Conclusion}\label{sec:jfm2:conclusion}

The lift and drag forces acting on a spherical particle in a single wall-bounded flow field are examined via numerical computation. Forces are obtained under the conditions of finite slip in both quiescent and linear shear flows. The effect of slip velocity, shear rate and wall separation are investigated by varying the slip and shear Reynolds number over the range $Re_\text{slip}, Re_\gamma = 10^{-3} - 10^{-1}$, and the wall separation distance over ${l^{\ast}} = 1.2 - 9.5$. The Navier-Stokes equations are solved using a finite-volume solver to find the fluid flow around the particle in large computational flow domains.

Based on the numerical results, we present a new lift force correlation in terms of three force coefficients, valid for any particle-wall separation distance (excluding contact), and $Re_\text{slip}, Re_\gamma \leq 0.1$. These three lift coefficients, $C_\text{L,1}$, $C_\text{L,2}$ and $C_\text{L,3}$, are defined to be functions of only shear rate, slip velocity and shear rate and only slip velocity, respectively, in addition to wall distance. First a correlation for the slip based lift coefficient ($C_\text{L,3}$) is proposed based on the lift results obtained for a particle translating in quiescent flow. This coefficient, which is independent of shear, reduces to the unbounded value of zero in the limit ${l^{\ast}} \rightarrow \infty$, and asymptotes to the inner-region theoretical value in the limits of $Re \rightarrow 0$ and ${l^{\ast}} \rightarrow 1$. The shear based lift coefficient, $C_\text{L,1}$ is adopted from our previous study \citep{EkanayakeJFM1}. The remaining coefficient, $C_\text{L,2}$, is calculated by subtracting the force contributions due to pure shear ($C_\text{L,1}$) and pure slip ($C_\text{L,3}$) from the numerical lift force results that are computed in a linear flow for both leading and lagging freely-rotating particles. By combining existing inner and outer-region based lift correlations, a new expression is then proposed to capture the $C_\text{L,2}$ coefficient behaviour. The net lift model obtained by combing all three lift coefficients covers both strong slip and shear flows and is applicable for negative or positive slip and shear rates. To our knowledge, this is the first lift model proposed for a rigid particle that accurately captures the transition in lift force behaviour between the inner and outer regions.

The performance of existing drag models are also compared against the numerical drag results for $Re_\text{slip}, Re_\gamma \leq 0.1$. The drag force coefficients computed for both non-rotating and freely-rotating particles in quiescent flows agree reasonably well with the inner-region \citet{Faxen22} predictions over the entire wall separation range. For linear shear flows, the most accurate drag predictions are obtained using the \citet{EkanayakeJFM1} and \citet{Faxen22} drag models.

The behaviour of freely translating neutrally-buoyant particles in a linear shear flow and sedimenting particles in both quiescent and linear shear flows are examined using the new lift correlations and examined drag correlations. The results are validated against numerical \citep{EkanayakeJFM1} and experimental values \citep{Takemura,TakemuraMagnaudet2}. The new lift correlation captures the numerical lift coefficient variation of the freely-translating neutrally-buoyant particles reasonably well. The correlation predicts the shear dependency of the lift coefficient and reduces to zero as the separation distance increases. The computed migration values for buoyant particles, using the new lift correlation also agree well with the experimental measurements in quiescent flows. While the inner-region slip based drag coefficient given by Faxen performs well for small $Re_\text{slip}<0.1$, the inner-outer-region-based correlation by \citet{Takemura} captures the drag variation when $0.1<Re_\text{slip}<1$. For buoyant particles in linear shear flows, the proposed lift model performs better than the other existing theoretical models when predicting the migration velocity for both positive and negative shear rates. However, a significant difference, noted at the lowest slip Reynolds number may be attributed to the measurements' uncertainties mentioned in the experimental study.

Overall, the proposed new lift correlations, valid for any particle-wall separation distance, will aid in providing accurate constitutive equations for interphase forces and will provide new opportunities to simulate many critical multiphase biological and industrial problems.

\section{Acknowledgements}
Support from the Australian Research Council (LP160100786) and CSL is gratefully acknowledged. One of the authors (NE) acknowledges the support from the Melbourne Research Scholarships program of The Melbourne University. Declaration of Interests: The authors report no conflict of interest.

\appendix
\section{}\label{appA}
Summary sheet of the recommended equations for lift and drag forces for a freely-rotating particle based on the analysis in this paper.

\subsection{Lift force}
\begin{gather}
    F^{\ast}_\text{L}=C_\text{L,1}Re_\gamma^2 + \text{sgn}({\gamma}^{\ast})C_\text{L,2}Re_\text{slip}Re_\gamma + C_\text{L,3}Re_\text{slip}^2 \quad\quad     \tag{\ref{eq:jfm2:netlift_allregion_0}}\notag\\
    F^{\ast}_\text{L} = F_\text{L} \rho/\mu^2\notag\\
    {\gamma}^{\ast} = \gamma a/u_\text{slip} \notag\\
    Re_\text{slip} = |a\rho u_\text{slip}/\mu|\notag\\
    Re_\gamma = |\gamma a^2\rho/\mu|
    \notag\\
    \hline\notag
    \intertext{Correlations from \citep{EkanayakeJFM1}}\notag
    {C_\text{L,1}} = f_\text{1}(Re_\gamma){C^\text{wb,out}_\text{L,1}}(l^{\ast}/{L_\text{G}}^{\ast}) + f_\text{2}(Re_\gamma){C^\text{wb,in$^\prime$}_\text{L,1}}({1/l^{\ast}})\notag\tag{3.3}\\
    f_\text{1}(Re_\gamma) = 0.9250\exp{(-0.3500 Re_\gamma)}-0.0135\exp{(-7000 Re_\gamma)};\notag\tag{3.4}\\
    {C^\text{wb,out}_\text{L,1}}(l^{\ast}/L_\text{G}^{\ast}) = 1.982\exp\big[{-0.1150\big({l^{\ast}}/{L_\text{G}^{\ast}}\big)^2-0.2771\big({l^{\ast}}/{L_\text{G}^{\ast}}\big)}\big];\notag\tag{3.5}\\
    f_\text{2}(Re_\gamma) =1+\sqrt{Re_\gamma};\notag\tag{3.6}\\
    C^\text{wb,in$^\prime$}_\text{L,1}({1/l^{\ast}})=1.0575(1/{l^{\ast}})-2.4007(1/{l^{\ast}})^{2}-1.9610(1/{l^{\ast}})^{3}
    \notag{\tag{3.3}}^\dagger\\\notag\\\hline\notag\\
    C_\text{L,3}=C_\text{L,3}^\text{wb,out}(l^{\ast}/{L_\text{S}}^{\ast})  + {C^\text{wb,in$^\prime$}_\text{L,3}}({1/l^{\ast}})\notag\tag{\ref{eq:lift0shear_inout}}\\
    C_\text{L,3}^\text{wb,out} = {6\pi}\big[3/\big(32+2\big({l}/{L_\text{S}}\big)+3.8({l}/{L_\text{S}})^2+0.049({l}/{L_\text{S}})^3)\big]\notag\tag{\ref{TM2004}}\\
    {C^\text{wb,in$^\prime$}_\text{L,3}}=0.4757({1}/{l^{\ast}}) -1.268({1}/{l^{\ast}})^2 + 0.683({1}/{l^{\ast}})^3
    \notag{\tag{\ref{eq:jfm2:lift0shear}}}^\dagger\\\notag\\
    \hline\notag\\
    C_\text{L,2}=-{C_\text{L,2}^\text{wb,out}}\frac{1}{\lvert{\gamma}^{\ast}\rvert} + {C_\text{L,2}^\text{wb,in$^\prime$}}({1/l^{\ast}})\notag\tag{\ref{eq:cl2_in-out}}\\
    {C_\text{L,2}^\text{wb,out}} = f(\epsilon, l/L_\text{G})C_\text{L,2}^\text{ub}\notag\tag{\ref{eq:slipshearlift-f}}\\
    C_\text{L,2}^\text{ub} = \frac{9}{\pi} \epsilon J(\epsilon)\notag\tag{\ref{eq:unbounded_lift}}\\
    f(\epsilon, l/L_\text{G}) = 1-\text{exp}\Big[-\dfrac{11}{96}\pi^2(l/L_\text{G})/J(\epsilon)\Big]\notag\tag{\ref{f_TM}}\\
    J(\epsilon) = 2.255(1+0.20\epsilon^{-2})^{-3/2}\notag\tag{\ref{Leg}}\\
    {C_\text{L,2}^\text{wb,in$^\prime$}}=  -2.6729-0.8373({1}/{l^{\ast}})+0.4683({1}/{l^{\ast}})^{2}\notag{\tag{\ref{CM2}}}^\dagger
    \\\notag\\
    \hline\notag
\end{gather}
\footnotetext{$^\dagger$ Correspondence to referred equation, but without the highest order term of with respect to ($1/l^{\ast}$)}
\subsection{Drag force}
\begin{gather}
    {F_\text{D}^{\ast}} = -\text{sgn}(\gamma)C_\text{D,1}Re_{\gamma} -\text{sgn}(u_\text{slip})C_\text{D,2}Re_{\text{slip}}\notag\tag{\ref{eq:jfm2:netdrag_allregion_0}}\\
    C_\text{D,2} = C_\text{D,2}^\text{ub}+C_\text{D,2}^\text{wb,in} \notag\tag{\ref{eq:drag_wb_faxen_cof}}\\
    C_\text{D,2}^\text{ub} =  6\pi \notag\tag{\ref{IC_Clift}}\\
    C_\text{D,2}^\text{wb,in} = 6\pi\Big(\Big[1-{9}/{16}(1/l^{\ast})+{1}/{8}(1/l^{\ast})^3-{45}/{256}(1/l^{\ast})^4\notag\\
    -{1}/{16}(1/l^{\ast})^5\Big]^{-1} -1\Big)\notag\tag{\ref{eq:drag_wb_faxen2}}\\
    C_\text{D,1} = ({15\pi}/{8})(1/l^{\ast})^2\Big[1+{9}/{16}(1/l^{\ast})+0.5801(1/l^{\ast})^2-3.34(1/l^{\ast})^3 \notag\\
    +4.15(1/l^{\ast})^4\Big]+(3.001Re_\gamma^2 -1.025Re_\gamma) \notag\tag{\ref{eq:drag_cd1_nilanka}}\\
\end{gather}
\bibliographystyle{jfm}
\bibliography{ekanayake20b}

\end{document}